\documentstyle[12pt,aasms4,amssym,psfig]{article}

\begin{document}

\bibliographystyle{apjproc}
\bibliographystyle{abbrv}

\title{Spectral Evolution of The Parsec-Scale Jet in The Quasar 3C\,345}
\author{ Andrew P. Lobanov \altaffilmark{1,2}
 and J. Anton Zensus\altaffilmark{1,2}}

\altaffiltext{1}{Max--Planck--Institut f\"ur Radioastronomie, Auf dem
H\"ugel 69, Bonn 53121, Germany. E--mail: alobanov@mpifr-bonn.mpg.de,
azensus@mpifr-bonn.mpg.de}
\altaffiltext{2}{National Radio Astronomy Observatory, 520 Edgemont Rd., 
Charlottesville, VA 22903, USA}

\begin{abstract}

The long--term evolution of the synchrotron emission from the
parsec--scale jet in the quasar 3C\,345 is analysed, on the basis of
multi--frequency monitoring with very long baseline interferometry
(VLBI) and covering the period 1979--1994.  We demonstrate that the
compact radio structure of 3C\,345 can be adequately represented by
Gaussian model fits and that the model fits at different frequencies
are sufficiently reliable for studying the spectral properties of the
jet. We combine the model fits from 44 VLBI observations of 3C\,345
made at 8 different frequencies between 2.3 and 100\,GHz.  This
combined database is used for deriving the basic properties of the
synchrotron spectra of the VLBI core and the moving features observed
in the jet. We calculate the turnover frequency, turnover flux
density, integrated 4--25\,GHz flux and 4--25\,GHz luminosity of the
core and the moving features. The core has an estimated mean
luminosity $L_{\rm core} = (7.1\pm 3.5) \cdot
10^{42}h^{-2}$\,erg\,s$^{-1}$; the estimated total luminosity of 3C345
on parsec scales is $\approx 3
\cdot 10^{43}h^{-2}$\,erg\,s$^{-1}$ (about 1\% of the observed
luminosity of the source between the radio to infrared regimes).  The
luminosities of the core and most of the moving features decrease at
the average rate of $1.2\cdot 10^{35}h^{-2}\, (0.74\pm
0.06)^{t-1979.0}$\,erg\,s$^{-2}$ ($t$ measured in years).  The derived
luminosity variations require intrinsic acceleration of the moving
features.  The turnover frequency of one of the moving features
reaches a peak during the above period.  The combination of the
overall spectral and kinematic changes in that feature cannot be
reproduced satisfactorily by relativistic shocks, which may indicate
rapid dissipation in shocks.  The spectral changes in the core can be
reconciled with a shock or dense plasma condensation traveling through
the region where the jet becomes optically thin. We are able to
describe the evolution of the core spectrum by a sequence of 5
flare--like events characterized by an exponential rise and decay of
the particle number density of the material injected into the jet. The
same model is also capable of predicting the changes in the flux
density observed in the core.  The flares occur approximately every
3.5--4 years, roughly correlating with appearances of new moving
features in the jet, and indicating that a quasi--periodic process in
the nucleus may be driving the observed emission and structural
evolution of 3C\,345.

\end{abstract}


\newpage
\section {Introduction \label{intro}}

The 16$^{\rm m}$ quasar 3C\,345 ($z=0.595$, Hewitt \& Burbidge 1993)
is one of the best examples of an active galactic nucleus (AGN)
showing structural and flux variability on parsec scales around a
compact unresolved radio core (for comprehensive reviews, see Zensus,
Krichbaum, \& Lobanov 1995, and Zensus 1997).  The source has been
detected in all wave bands except for $\gamma$--ray ($E>100$\,MeV)
where only an upper limit of the flux is known \cite{fbc+94}. In the
X-ray regime, the source is weak and possibly variable (Halpern 1982;
Makino 1989; Worrall and Wilkes 1990).  At 1\,keV, the flux density is
$S_{\rm x}=0.39\pm$0.03\,mJy. The spectral index is $\alpha_{\rm
x}\approx -0.9$ in the 0.2--2.0\,keV spectral band \cite{unw+94}.  In
the ultraviolet, the flux density is $S_{\rm uv} \approx 6$\,mJy,
assuming a column density of neutral hydrogen $N_{\rm
H}=10^{20}$\,cm$^{-2}$ and a spectral slope similar to that of the
X--ray emission
\cite{mma+94}.  In the optical regime, the source is highly polarized
\cite{ms84} and variable.  The observed variations are possibly
quasi--periodic with a period of $\approx$1560\,days (Babadzhanyants
and Belokon' 1984; Kidger 1990), although it has been suggested that
the light curve may originate from a non--linear and non--stationary
stochastic process \cite{vio+91}.  Infrared observations of 3C\,345
show $S_{\rm 60\mu m}\approx 0.7$\,Jy \cite{in88}.

The total radio flux density of 3C\,345 has been monitored at 5, 8,
and 15\,GHz \cite{all+96}, and at 22 and 37\,GHz \cite{ter+98}. The
source has also been monitored with the Green Bank Interferometer at
2.7 and 8.1\,GHz \cite{wal+91}. The continuum radio spectrum is flat
up to 10\,GHz, and steepens towards higher frequencies, with the
spectral index ranging from $-$0.9 to $-$1.4. \cite{bre+86}.  The time
scale of the variability shortens towards higher frequencies,
suggesting the presence of at least two emitting regions in the source
core
\cite{bre+86}.

VLA observations of 3C\,345 \cite{kwr89} show a faint halo around a
bright core, and an extended kiloparsec--scale jet.  On smaller
scales, the source structure is predominantly of core--jet type.  The
parsec--scale emission of 3C\,345 has been monitored extensively with
very long baseline interferometry (VLBI) (Unwin et al.,\ 1983;
Biretta, Moore \& Cohen 1986, hereafter BMC86; Zensus, Cohen, \& Unwin
1995, hereafter ZCU95).  The monitoring has been performed at 5, 8.4,
10.7, and 22.3\,GHz, and additional observations have been made at
2.3, 43, 89, and 100\,GHz.  Recent VLBI observations covering the
epochs between 1989 and 1997 are presented in Lobanov (1996, hereafter
L96) and Ros et. al (1999).

The relativistic jet model \cite{br78} has been applied to 3C\,345 to
explain the nature of the enhanced emission regions in the jet.  Using
the X-ray data to constrain the jet kinematics, ZCU95 and Unwin et
al. (1994) have derived the physical conditions of the jet from a
model that combines the inhomogeneous--jet model of K\"onigl (1981)
for the core with homogeneous synchrotron spheres for the jet
components \cite{coh85}.  Unwin et al. (1997) have analysed a
correlation between the X-ray variability and parsec--scale radio
structure of 3C\,345.  Steffen et al. (1995) derived a helical jet
model to explain the observed component trajectories and flux density
variations. Raba\c{c}a and Zensus (1994) showed that the flux
evolution of the jet component C4 is consistent with a strong shock
concentrated in a narrow region.  Wardle et al. (1994) have shown that the
polarization structure can also be reproduced within the relativistic
shock model.

Because of the limitations of VLBI observations, spectral properties
and spectral variations of parsec--scale emission have so far not been
studied in detail. Such a study requires a homogeneous VLB array,
reliable amplitude calibrations, quasi--simultaneous observations at
different frequencies, and analysing the emission at different
frequencies continuously along the jet. For two recent epochs, the
turnover frequency distributions in 3C\,345 and 3C\,273 have been
mapped and the magnetic field profiles have been obtained in (Lobanov
\& Zensus 1994, Lobanov, Carrara, \& Zensus 1997, Lobanov 1998b). Owing
to the insufficient data quality, it is not possible to extend this
analysis to earlier observations; we must therefore resort to a simpler
approach, combining observations made at nearby epochs, and
using model fits to represent the source structure. At this time, this
is the only plausible means to study, albeit crudely, the long--term
spectral evolution of parsec--scale radio emission.

In this paper, we use the available multi--frequency VLBI results to
study the spectral evolution of different regions in the jet of
3C\,345.  In section \ref{sc:data}, we give a summary of the VLBI
data, and describe the procedures used for data analysis.  We
calculate the parameters of simple synchrotron spectra: the turnover
frequency $\nu_{\rm m}$, the turnover flux density $S_{\rm m}$, and
the integrated 4--25\,GHz flux $S_{\rm int}$.  The evolution of the
spectrum and luminosity evolution of the jet are discussed in section
\ref{sc:spectrum}. In section \ref{sc:results}, the observed spectral
changes in the moving features of the jet are compared with the
predictions of the shock--in--jet model. In section~\ref{sc:konigl},
the observed variations of the spectrum and flux density of the VLBI core
are modeled by flare--like events developing in an inhomogeneous jet.

Throughout this paper, we use the positive definition of spectral
index, $\alpha$ ($S \propto \nu^\alpha$), a Hubble constant, $H_0 =
100\,h\, {\rm km\,s^{-1}\,Mpc^{-1}}$, and a deceleration parameter,
$q_0 = 0.5$.

\section {Multi--frequency VLBI data \label{sc:data}}

We use data from 44 VLBI observations of 3C\,345 made at 2.3, 5, 8.4,
10.7, 22.3, 43.2, 89.2, and 100\,GHz during the period
1979.25--1993.88 (see Table~\ref{tb-dat} for references).  Standard
VLBI analysis was applied to all the data, with CLEAN
$\delta$-components used for representing the source structure (see
Zensus, Diamond, \& Napier 1995 for the most recent treatment of the
subject). It is, however, very difficult to employ the CLEAN components
for describing the kinematic and emission properties of VLBI-scale
regions. We have therefore resorted to working with simple models of
Gaussian or spherical components fitted directly to the interferometric
visibility data.  The models for earlier epochs (1979--1984) were
taken from the literature; the models for later epochs have been made
from the original data. We used the MODELFIT routine of the Caltech VLBI
package (Pearson 1991) and fitted the complex baseline visibilities using
the maximum likelihood method (minimizing $\chi^2$, the weighted sum of
squares of the deviations between the data and the model). We evaluated
the goodness of the fit by the reduced chi-square, $\chi^2/(N_{\rm
datapoints}-N_{\rm parameters})$, and determined the uncertainties of
the fitted parameters from the corresponding confidence limits in the
$\chi^2$ distribution. This allowed us to account for increased
uncertainties in cases when some of the fitted parameters could not be
considered strictly independent (for instance, if the separation between
two Gaussian components is smaller than their sizes; see Pearson 1995
for an extensive discussion of the model fitting technique). This model
fitting technique has been shown to be a reliable tool for
representing the VLBI--scale structure and emission in many sources,
including 3C\,345 itself (ZCU95) and even more complex sources, such
as Cygnus~A \cite{kri+98}.

\placefigure{fg:36nov93map}

\subsection {Model fitting the source structure in 3C\,345 
\label{sc:modelfit}}

A VLBI image of 3C\,345 at 8.4\,GHz obtained from an observation made
in November 1993 is shown on the top in Figure~\ref{fg:36nov93map}.
The bottom image in Figure~\ref{fg:36nov93map} is obtained by modeling
the visibility data with 8 elliptical Gaussian components (the 8th
component corresponds to a weak, distant feature C1 lying outside the
plotted ranges). One can see the remarkable similarity between the
VLBI image and the model fit representation of the source
structure. This is also illustrated in Figure~\ref{fg:visfit}, where
we compare the representations of the visibility amplitudes and
closure phases by the model from CLEAN components and that from
Gaussian components. The two representations are barely
distinguishable only by the more rapid variations visible in the CLEAN
component model (because of a wider area of the image plane used for
determining the CLEAN component model). It is also remarkable that the
exclusion of either C6 or C7 from the Gaussian model deteriorates the
fit so much that it becomes impossible to achieve a satisfactory
agreement with the data, no matter what adjustments are made to the
remaining components. The presence of C6 and C7 in the jet is also
self--evident in the VLBI image of 3C\,345 made at a close epoch at
22\,GHz, with a resolution about 3 times better (L96). This and numerous
other examples of model fitting 3C\,345 convince us of the suitability
of Gaussian component models for describing the structure and emission
in 3C\,345.
 
In 3C\,345, the VLBI core, D, is probably stationary
\cite{tan+90}, and the bright components C2--C7 embedded in the jet
are receding from the core at apparent speeds ranging from 1 to 20$c$
(ZCU95, L96). We limit our present discussion to the core D and the
components C2--C5 which were dominating the source emission in the
period 1979--1993.  We obtain spectra and luminosities for D, C5, and
C4, and only make luminosity estimates for C2 and C3.  Owing to the
decreasing surface brightness of the extended jet, some of the recent
measurements for the oldest feature C2 are not reliable.  At distances
larger than $\approx 5$\,mas, the emission regions in the jet tend to
become more complicated, and generally more difficult to describe by a
single Gaussian component. However, the bulk of the data that we use
for our spectral fitting is related to the features of the jet at distances
smaller than 5\,mas, where the Gaussian representation is sufficient.

\placefigure{fg:visfit}

\subsection{Multi--frequency datasets}

The VLBI observations at different frequencies were typically not made
simultaneously; and we combine only those observing epochs which are
sufficiently close in time, so that the source variability will not
undermine the results.  The light curves for the total and component
flux densities yield time intervals of up to 6 months, during which the
changes do not exceed the typical errors inherited from the model
fitting.  We use this limit to combine the flux data into
multi--frequency datasets. A stricter limitation would result in too
few possibilities of constructing such sets.

The multi--frequency datasets are listed in Table~\ref{tb-dat}.  Each
set, except for the first two, contains observations at three or more
frequencies.  In all datasets, the lowest frequency is 5\,GHz or
lower.  If a component is detected only at two frequencies within a
given set, only the limiting values can be determined for the fitted
spectral parameters.

\placetable{tb-dat}

\subsection{Spectral fitting}

We assume that spectrum of a jet component is similar to that of a
homogeneous relativistic plasma with a power--law electron energy
distribution
\begin{equation}
S_{\nu} \propto \left( \frac{\nu}{\nu_1}
\right)^{5/2} \left\{ 1 - \exp \left[ - \left(
\frac{\nu_1}{\nu} \right)^{5/2-\alpha} \right]\right\} \label{eq:synspec}
\end{equation}
\cite{pac70}, where $\nu_1$ is the frequency at which the optical
depth is $\tau=1$, and $\alpha$ is the spectral index.  Instead of
$\nu_1$, we use the frequency of spectral maximum $\nu_{\rm m}$ (the
turnover frequency). The two can be related, using a simple
approximation $\nu_1 \approx 0.912 (-\alpha)^{0.386} \nu_{\rm m}$.
Since VLBI observations cover a relatively narrow spectral interval,
$\nu_{\rm m}$ may lie outside the observed frequency range.  For
such ill--constrained cases, we first fit the spectral data by
polynomial functions, and then analyse local curvature of the fits obtained.
We compare the value of the curvature obtained with the
theoretical curvature of the synchrotron spectrum, and make an
estimate (often only an upper limit) of $\nu_{\rm m}$. A detailed
discussion of the procedure is given in Lobanov (1998b). The main steps
of spectral fitting can be summarized as follows:

1)~We make an approximate estimate of relevant frequency range,
($\nu_{\rm L}$, $\nu_{\rm H}$), outside which the flux density level
is negligibly small. For that purpose, we use the average measured
turnover frequency and the spectral spectral index in the compact
source. On the basis of these values, we estimate the frequencies
$\nu_{\rm L}$ and $\nu_{\rm H}$ at which the corresponding flux
density is at an arbitrarily low level (we use $S_{\rm cutoff} =
0.1$\,mJy). Let us stress here that no specific physical meaning is
attached to these quantities.  We add these spectral points to the
measured data solely for the purpose of ensuring a negative curvature
of the fitted polynomial curves, as required by the theoretical
spectral form (\ref{eq:synspec}). To account for possible
uncertainties in the estimated values, we allow 15\% variations of
$\nu_{\rm L}$ and $\nu_{\rm H}$, during each iterative cycle of the
fitting routine described below.

In our estimates of $\nu_{\rm L}$ and $\nu_{\rm H}$, we profit from the
sharpness of the transition between the high--frequency ($\nu > \nu_1$) and
low--frequency ($\nu < \nu_1$) spectral regimes determined by equation
\ref{eq:synspec}.  The resulting spectrum is determined by the
synchrotron self--absorption at frequencies $\nu\ll\nu_1$, and by the
electron energy distribution at $\nu\gg\nu_1$. Estimates of the
typical turnover frequency and energy spectral index in the jet of
3C345 based on our own calculations and on the results from Raba\c{c}a
and Zensus (1994) give $\nu_1 \approx 10$\,GHz and $\alpha\approx -0.7$.
Using these values, we calculate the low--frequency, $\nu_{\rm L}\sim
1$\,MHz, and high--frequency, $\nu_{\rm H}\sim 1000$\,GHz, fiducial
spectral points at which the flux density level is about 0.1\,mJy, and
use these values as the initial guesses for the spectral fitting.  

2)~We fit, iteratively, the combined spectrum by polynomial functions,
allowing for limited variations of $\nu_{\rm L}$ and $\nu_{\rm H}$,
and aiming at achieving the best fit to the measured data points.
From the fits, we obtain the initial estimates of the basic parameters
of the spectra: the turnover frequency, $\nu_{\rm m}$, turnover flux
density, $S_{\rm m}$, and the integrated flux, $S_{\rm int}$ (the
integration limits are set to a range of 4--25\,GHz, the typical range
of the observing frequencies).

3.~The initial estimates of $\nu_{\rm m}$ are corrected and classified
as values or limits, by considering the local curvature of the fitted
polynomial forms, and deriving the ranges of reliable corrections from
the mean flux density errors in the multi--frequency datasets.  We
calculate the local curvature of the fitted spectra within the range
of observing frequencies. The derived curvature can be compared with
the values obtained from analytical or numerical calculations of the
synchrotron spectrum. In this paper, we derive the corrected value of
the turnover frequency, by equating the fitted and the theoretical
spectral curvatures.

4.~We then fit the data with the synchrotron spectral form described
by (\ref{eq:synspec}), using the corrected value of the turnover
frequency, and obtaining the final estimates of $S_{\rm m}$ and
$S_{\rm int}$ (the differences between the initial and final estimates
of $S_{\rm int}$ are insignificant, since, by virtue of the 
fitting procedure applied, both the polynomial and synchrotron spectral forms
are optimized to provide the best fit within the range of observing
frequencies, which covers, in most cases, the range of 4--25\,GHz used for
calculating $S_{\rm m}$). We obtain the formal errors of the fitted
spectral parameters from Monte Carlo simulations, assuming Gaussian
distribution of VLBI flux density errors, and analysing the
distributions of the $\chi^2$ parameters of the fits to the simulated
datasets.

\placefigure{fg:kappa}

\subsubsection{Curvature of the fits \label{sc:frcoverage}}

Here, we outline the application of the local curvature of the fitted
polynomial forms to correcting and classifying the initial estimates
of the spectral parameters.  Following Apostol (1969), the local
curvature, $\kappa$, of spectral fits, $S(\nu)$, can be written as:
\begin{equation}
\label{eq:korn1}
\kappa = \frac{{\rm d}^2 S}{{\rm d}\nu^2} \left[1+\left(\frac{{\rm d}S}{{\rm d}\nu}\right)^2
\right]^{-1/3}\, .
\end{equation}
If the spectral index, $\alpha_{\rm o}$, and the local curvature,
$\kappa_{\rm o}$, of a polynomial fit are determined at a frequency
$\nu_{\rm o}$, then the turnover frequency, $\nu_{\rm m}$, can be
estimated from the adopted theoretical synchrotron spectrum. Using the
derived $\alpha_{\rm o}$ and $\kappa_{\rm o}$, we determine the ratio
$\xi(\alpha_{\rm o}, \kappa_{\rm o}) = \nu_{\rm m}/\nu_{\rm o}$, from
the adopted spectral form described by (\ref{eq:synspec}).  The
corresponding turnover frequency is then
\begin{equation}
\label{eq:curv1}
\nu_{\rm m} = \nu_{\rm o} \xi(\alpha_{\rm o}, \kappa_{\rm o})\, .
\end{equation}
Figure~\ref{fg:kappa} relates the curvature $\kappa$ of the
homogeneous synchrotron spectrum to the ratio $\xi(\alpha_{\rm o},
\kappa_{\rm o})$. For frequencies increasingly deviating from the
turnover frequency, $\kappa$
becomes progressively smaller, thereby limiting the ranges of
applicability of the corrections described by (\ref{eq:curv1}). For
data covering the frequencies from $\nu_{\rm l}$ to $\nu_{\rm h}$, and
for negligible ($\le 1\%$) flux density errors, we expect the
corrections to give reliable estimates for the turnover frequencies
lying within the range of $0.05\nu_{\rm l} <
\nu_{\rm m} < 2\nu_{\rm h}$. For a spectral dataset containing 
$N$ measured flux densities, $S_{\nu,i}$, and flux density errors,
$\sigma_{i}$, we can estimate the minimum measurable curvature,
$\kappa_{\rm min}$, warranted by the data. We use the average
fractional flux density error, $\sigma_{\rm S,N} =
\langle\sigma_{i}/S_{\nu,i}\rangle$, and obtain
\begin{equation}
\label{eq:curverror}
\kappa_{\rm min} = 2 \hat{\sigma} [1 + (\alpha + 2 \hat{\sigma} 
\nu_{\rm h,l})^2]^{-1/3}\, ,
\end{equation}
where $\hat{\sigma} = \log(1-\sigma_{\rm S,N})$ and $\nu_{\rm h,l} =
\log(\nu_{\rm h})/\log(\nu_{\rm l})$. In Figure~\ref{fg:kappa}, we mark
the ranges of $\xi(\alpha, \kappa)$ corresponding to $\kappa_{\rm
min}$ estimated for spectral datasets with average fractional flux
density errors $\sigma_{\rm S,N} =$1, 5, 10, and 20\%. The
$\sigma_{\rm S,N}$ of our spectral datasets are given in the 4$^{\rm
th}$ column of Table~\ref{tb-fits}. The majority of the calculated
$\sigma_{\rm S,N}$ are smaller than 0.1, which implies that in most cases
the curvature corrections give reliable results for the range of
frequencies $0.3\nu_{\rm l} < \nu < 1.3\nu_{\rm h}$, considering a
typical synchrotron spectrum with $\alpha =-0.5$.  We use the measured
$\sigma_{\rm S,N}$ and the corresponding ranges of
$\xi(\alpha,\kappa)$ for calculating the final estimates of the
spectral parameters and deciding whether these estimates should be
treated as the values of, or only the limits on the respective quantities.

\placetable{tb-fits}

\section{Properties of the synchrotron spectra of the jet components 
\label{sc:spectrum}}

Table~\ref{tb-fits} lists the derived values and limits (shown in
brackets) of the synchrotron spectra in the VLBI core and jet
components C2, C3, C4, and C5.  For each feature and epoch, the table
also shows the spectral index, $\alpha_{\rm fit}$ of the fitted
synchrotron spectra, and gives the limits on the bulk Lorentz
factors and luminosities of the features, which will be discussed in
section~\ref{sc:lum-kin}. Figure~\ref{fg:c5fits} illustrates the fitting
results for the component C5. The dotted lines represent third order
polynomial fits to the data. The solid lines show the synchrotron
spectra derived after the application of the local curvature corrections 
described in section~\ref{sc:frcoverage}. If
the spectral turnover is covered by the data (epochs 1987.3, 1988.2 in
Figure~\ref{fg:c5fits}), the curvature corrections were applied at the
turnover point obtained from the polynomial fit. In other cases, the
corrections were applied at the lowest observed frequency. For the
datasets containing only two spectral points, only crude estimates of upper
limits of $\nu_{\rm m}$ and lower limits of $S_{\rm m}$ were made
(these estimates are largely determined by the typical $\nu_{\rm L}$
and $\nu_{\rm H}$ inferred from the fits to other datasets for C5).  We
will base our discussion of the spectral evolution on the derived
spectral properties of the core and C5 for which the spectral data are
most complete (we will also make some remarks on C4 for which we have only 4
epochs with estimated values of $\nu_{\rm m}$ and $S_{\rm m}$.  For
C2, C3, and C4, largely represented by the limiting values, only the
luminosity estimates will be discussed.

\placefigure{fg:c5fits}
 
\subsection{Evolution of the turnover frequency}

Figures~\ref{fg:newevol}--\ref{fg:turnflux} present the fitted
turnover frequencies and flux densities for C4, C5, and D. C4 is
represented mostly by the limiting values of $S_{\rm m}$ and $\nu_{\rm
m}$.  The data for C5 are more complete. In the core, the highest
values of $\nu_{\rm m}$ ($>20$\,GHz) may be underestimated, owing to the
uncertainty in determining the spectral index in cases when only the
optically thick part of spectrum is sampled. The discrepancy between
the derived values and lower limits of $S_{\rm m}$ in C4 indicates
that the fits to the two--point spectral datasets may suffer from
systematic errors. 

The increases of the turnover frequency of the core do not precede or
accompany the ejection of new jet components, but rather occur some
time after the ejections. At the epochs when the components were first
detected (1980.2: C4, 1983.1: C5---BMC86; 1989.2:
C6,C7---B{\aa}{\aa}th et al. 1992), the turnover frequency in the core
was comparatively low.  The changes of $S_{\rm m}$ in the core are
correlated with the flux density variations in the core and with the 
changes in total flux density at 22\,GHz (although the correlation with the
total flux density is weaker, owing to substantial contributions of the jet
components; L96).

The situation is different in C5. Between 1984 and 1990, the
turnover flux density of C5 remained within the range of 3.8--5.6\,Jy,
whereas the 5, 8, and 22\,GHz flux densities of the component were varying
between 1 and 6\,Jy (L96).  This suggests that the observed spectrum
could be significantly affected by Doppler boosting.
The turnover frequency of C5 shows a peak in time, with the maximum
occurring between 1985.8 and 1988.2. The variations of $\nu_{\rm m}$
may be similar in C4, if we consider the upper limits in 1981.8 and
1982.4. 

\placefigure{fg:newevol}
\placefigure{fg:turnflux}

\subsection{Luminosity and kinematics \label{sc:lum-kin}}

The luminosities of the core, and the jet components C2--C5 can be
derived from the measured $S_{\rm int}$ which does not require the
turnover frequency to be known. The kinematic parameters of the jet
components can be inferred from the polynomial fits to their observed
component trajectories (L96). Since $S_{\rm int}$ is an integral
quantity, the anisotropy of relativistically beamed emission cannot be
accounted for by the commonly used $\delta^{2-\alpha}$ correction,
because the turnover point is often located within the 4--25\,GHz frequency
limits, and the spectral index changes within the range of integration.
To account for the anisotropy of the emission, we calculate the
fraction, $\Omega/4\pi$, determined by the solid angle, $\Omega$, which
contains most of the beamed emission.  We use the measured apparent
velocity, $\beta_{\rm app}$ and determine the minimum Lorentz factor
$\gamma_{\rm min}=(1+\beta_{\rm app}^2)^{1/2}$ that can reproduce the
observed $\beta_{\rm app}$. Under this condition, the component
kinematics are defined solely by $\gamma_{\rm min}$, since $\sin\theta
= 1/\gamma_{\rm min}$ and $\delta = \gamma_{\rm min}$, where $\theta$
is the angle between the velocity vector of the component and the line of
sight. This gives an upper limit for the 4--25\,GHz
luminosity
\begin{equation}
\label{zero}
L_{\rm 4-25GHz} \approx \frac{8\pi c^2 S_{\rm int}}{H_0^2} 
\left[ 1 - \cos \left(\frac{1}{\gamma_{\rm min}}\right)\right] \left[
1+z - \sqrt{1+z}\right]^2\, .
\end{equation}

\placefigure{fg:evolbeta}

The derived luminosities and minimum Lorentz factors are given in
Table~\ref{tb-fits}. Figure~\ref{fg:evolbeta} shows the luminosity
changes in the core and the jet components.  We estimate $\gamma_{\rm
D} \approx 3.5$, assuming that the Lorentz factor of the core is
similar to the initial Lorentz factors of the moving components
(ZCU95).  This yields $L_{\rm D} = (7.1\pm 3.5) \cdot 10^{42}
h^{-2}$\,erg\,s$^{-1}$ for the average luminosity of the core. We
estimate the total 4--25\,GHz luminosity of the jet features as $L_{\rm
jet} \approx 3\cdot 10^{43} h^{-2}$\,erg\,s$^{-1}$, which amounts to
about 1\% of the observed luminosity of 3C345 in the radio to infrared
range \cite{bre+86}.  The average decrease rate for all components is
$\approx1.2
\cdot 10^{35} h^{-2}\,(0.74\pm 0.06)^{t - 1979.0}$\,erg\,s$^{-2}$ (we
use the 1981-88 data to determine the luminosity decrease in the
core).  The derived rates of decrease are given in
Table~\ref{tb-evolbet}.  Comparable rates of energy losses in
different jet components imply similar physical conditions.  The only
exception to the observed similarity is the luminosity of C4 which
decreases almost twice as fast.  This difference can be caused by
several factors:

\placetable{tb-evolbet}

1)~Because of the narrow observed frequency range, the derived
luminosities might not adequately reflect the physical
conditions in the jet, and the observed regularities might be
coincidental. However, most of the component spectra have their maxima
within or near the 4--25\,GHz range, so that their evolutions must be
similar in order to produce comparable rates of decrease of the luminosity.

2)~It is possible that we are underestimating the Lorentz factors of the
components C2, C3, and C5.  However, it would require very high
Lorentz factors ($\gamma_{\rm C3}\approx 35$, $\gamma_{\rm C5} \approx
26$), to match the slope of C4. Here, we have taken the luminosities of
C3 and C5 at their first epochs, and calculated their expected
luminosities at 1993.0, assuming that the components had the same
rate of decrease as that of C4. From the calculated luminosities, the
required Lorentz factors were determined, using equation \ref{zero}.
The derived $\gamma_{\rm C3}$ and $\gamma_{\rm C5}$ are considerably
higher than the theoretical limits on the bulk Lorentz factor expected
in a jet ($\gamma\approx 20$; Abramovicz 1992; Henri and Pelletier
1992).

3)~C4 might represent a composition of two possibly related features
in the jet. The observations made in 1984--86 indicate that C4 could
be a blend of two emitting regions (ZCU95). The compound nature of C4
is not obvious at later epochs, and we can suggest that a weaker
subcomponent could not be distinguished from the emission of a
stronger and faster main subcomponent (C6 and C7 may represent a
similar situation: the faster and younger C7 overtakes the older and
slower C6; Lobanov 1996). This might result in an increase of
the measured proper motion and, correspondingly, in higher values of
$\gamma_{\rm min}$.

\placefigure{fg:loom}

The implications of the faster decrease in luminosity observed in C4 can
be understood better, if we study the evolution in the luminosity along the
jet.  Figure~\ref{fg:loom}a presents the correlations between the derived
luminosities of the components and the apparent distance traveled along the
jet
\begin{equation}
\label{eq:Rapp}
 R_{\rm app}(t) = \int^t_0 \mu(t)\, {\rm d}t \, ,
\end{equation}
where $\mu(t)$ is the proper motion of the compoment, and $t=0$ refers to its
epoch of ejection. Figure~\ref{fg:loom}a shows that the
luminosities of C2, C3, and C5 depend on $R_{\rm app}$ in a similar
fashion, so that it is possible to find a general trend for the
decrease of luminosity in the jet on the scales of up to 8\,mas.  The
linear fit to this trend, represented by the solid line in
Figure~\ref{fg:loom}a, corresponds to $\log (L_{\rm C2,3,5}) = 42.91 -
0.29\,R_{\rm app}$. The same procedure applied to C4 yields $\log
(L_{\rm C4}) = 44.35 - 0.66\,R_{\rm app}$. All fits are described in
Table~\ref{tb-lum}.

\placetable{tb-lum}

We now investigate whether the luminosities of the jet components are
better correlated in the rest frame of the jet.  In order to study the
luminosity evolution in the rest frame of the jet, certain assumptions
must be made about the component kinematics. We consider the
luminosity evolution in a curved jet with constant and variable
component speeds.  In the rest frame of the jet, the jet component
travels the distance
\begin{equation}
\label{eq:Rj}
R_{\rm j}(t)=(1+z)^{-1} \int^t_0 \frac{\beta(t)} {1 - \beta(t)\cos\theta(t)}{\rm d}t \, .
\end{equation}
The time variations of the component speed $\beta$ and viewing angle
$\theta$ are determined by the measured $\beta_{\rm app}$ and the chosen
evolution of the Lorentz factor.  

The observed trajectories of the components cannot accommodate a
single value of Lorentz factor $\gamma_{\rm jet}$, for all jet
components, unless $\gamma_{\rm jet}>\ge20$ (L96).  We find that
adopting $\gamma_{\rm jet} =20$ does not improve the agreement between
C4 and other jet features. If the individual Lorentz factors of the
jet components are allowed to be different, the kinematic constraints
from the observed component trajectories require that $\gamma_{\rm
C5}\ge7$, $\gamma_{\rm C4}\ge15$, $\gamma_{\rm C3,C2}\ge20$ (L96).  We
adopt these constraints and plot the resulting rest frame luminosity
evolution in Figure~\ref{fg:loom}b.  The component luminosities in
Figure~\ref{fg:loom}b are not well correlated (see Table~\ref{tb-lum}
which lists the corresponding rates of decrease of the luminosities).
At comparable distances, the luminosities of different components can
differ by more than one order of magnitude.  The decrease in the
luminosity of C5 is also faster than in other components.  These
discrepancies suggest that the Lorentz factors of the components may
vary along the jet. To illustrate that, we consider a jet that carries
the least kinetic power, so that each component in the jet always
moves at its respective $\gamma_{\rm min}$, as determined from the
variations of the observed $\beta_{\rm app}$.  The resulting
$\gamma(R_{\rm j})$ tracks are shown in Figure~\ref{fg:loom}d. The
corresponding luminosities of the components are presented in
Figure~\ref{fg:loom}c. The solid line in Figure~\ref{fg:loom}c
represents overall decrease in luminosity $\log(L_{\rm jet})
\approx42.06 - 0.03\,R_{\rm j}$ derived for all the components. With
exception of C2 (for which the calculations of luminosity may be not
reliable), the individual evolutions of the components do not deviate
significantly from the overall decrease (see Table~\ref{tb-lum}).
Assuming that the mechanism of emission is the same in all of the
components in the jet, the observed similarity of the decrease in
luminosities of the components suggests that the Lorentz factor should
change along the jet, in order to explain the variations of these
luminosities . The intrinsic accelerations are also required from the
kinematic constraints in the jet (Lobanov \& Zensus 1996). Unwin et
al. (1997) find that the component C7 in the jet must accelerate from
$\gamma\approx5$ to $\gamma\approx10$ in order to reproduce the X--ray
variations observed in 3C\,345. The derived rate of acceleration of C7
is then consistent with the changes in $\gamma_{\rm min}$  in C3 and C4.

\section{Spectral evolution in the relativistic shock model\label{sc:results}}

Consider a relativistic shock propagating down a conical jet with a
constant half--opening angle $\phi$. The shock expands sideways
adiabatically, and maintains the condition of a plane shock, so that
its longitudinal dimension is much smaller than the transversal. A
power--law distribution of the energy of electrons, $N(\gamma)d\gamma
\propto \gamma^{-s} d\gamma$, is assumed.  With increasing distance,
$R$, from the origin of the jet, the magnetic field decreases as
$R^{-a}$, with $a=1$ for the field perpendicular to the jet axis, and
$a=2$ for the field parallel to the axis (intermediate values of $a$
are also possible).  The shock passes through three basic evolutionary
stages at which its emission is subsequently dominated by the Compton,
synchrotron and adiabatic losses (Marscher, Gear, \& Travis 1991).  In
the observer's frame, the received spectrum of the shock emission can
be written in the form:
\begin{equation}
\label{7}
S(\nu) \propto R^\xi\nu^{-\zeta}\delta^{(s+3)/2} \, ,
\end{equation}
where $\xi$ and $\zeta$ describe different stages of evolution of the
shock.  The turnover flux density is:
\begin{equation}
\label{8}
S_{\rm m} \propto B^{-1/2}\nu_{\rm m}^{5/2}R^2 \propto R^{(4+a)/2}\nu_{\rm m}^{5/2} 
\end{equation}
\cite{caw91}. We further assume that the Doppler factor
$\delta\propto R^b$, where $b$ is constant. 
 Combining (\ref{7}) and (\ref{8}), we obtain for the turnover frequency
\begin{equation}
\label{9}
\nu_{\rm m} \propto R^{[\xi-(4+a)/2+b(s+3)/2]/[5/2+\zeta]} \, .
\end{equation}
Using $\xi$ and $\zeta$ calculated by Marscher (1990), we write
expressions for $\nu_{\rm m}$ and $S_{\rm m}$ at each stage of the evolution
of the shock.
\newline
1)~Compton--loss stage\quad ($\xi=[(11-s)-a(s+1)]/8; \quad \zeta = s/2$)
\begin{equation}
\label{10}
\nu_{\rm m} \propto R^{-[(5+s)(a+1)-4b(s+3)]/[4(s+5)]} \, ,
\end{equation}
\begin{equation}
\label{10a}
S_{\rm m} \propto \nu_{\rm m}^{[(a-11)(s+5) - 20b(s+3)]/[2(s+5)(a+1) - 8b(s+3)]} \,.
\end{equation}
2)~Synchrotron--loss stage\quad ($\xi=-[4(s-1)+3a(s+1)]/6; \quad
\zeta = s/2$)
\begin{equation}
\label{11}
\nu_{\rm m} \propto R^{-[3a(s-1)+4(s+2)-3b(s+3)]/[3(s+5)]} \, ,
\end{equation}
\begin{equation}
\label{11a}
S_{\rm m} \propto \nu_{\rm m}^{[3b(3s+10)-(3a+2)(2s-5)]/[3b(s+3)-3a(s-1)-4(s+2)]}\, .
\end{equation}
3)~Adiabatic--loss stage \quad ($\xi=[2(5-2s)-3a(s+1)]/6; \quad
\zeta = (s-1)/2$)
\begin{equation}
\label{12}
\nu_{\rm m} \propto R^{-[2(2s+1)+3(a-b)(s+2)]/[3(s+4)]} \, ,
\end{equation}
\begin{equation}
\label{12a}
S_{\rm m} = \nu_{\rm m}^{[(19-4s)-3a(2s+3)+3b(3s+7)]/[3(b-a)(s+2) - 2(2s+1)]} \,.
\end{equation}
$R$ in formulae \ref{10}--\ref{12a} is still measured in the rest
frame of the jet, and should be transformed to the observer's
frame, using equation ({\ref{eq:Rj}). 

We now consider the observed changes of $S_{\rm m}$ and $\nu_{\rm m}$
in the jet of 3C\,345 in the framework of the dependencies
(\ref{10})--(\ref{12a}).  For simplicity, we assume $a\approx const$,
and $s\approx const$. Spectral changes in a shock can be then
described by three parameters: $\rho$ ($S_{\rm m} \propto \nu_{\rm
m}^{\rho}$), $\varepsilon$ ($\nu_{\rm m} \propto R^{\varepsilon}$),
and $b$ ($\delta \propto R^b$) which we will also be calling the $b$--parameter.
Figure~\ref{fg:shockpar} illustrates the relations between $\rho$,
$\varepsilon$, and $b$, for each evolutionary stage of the model. The
shown curves do not differ significantly, for conceivable ranges of
$a$ ($1<a<2$) and $s$ ($1<s<3$).

\placefigure{fg:shockpar}

\subsection{The VLBI core \label{sc:core}}

The spectral evolution of the core is shown in
Figure~\ref{fg:corevol}. We are unable to find suitable model
parameters for two intervals: 1982.4--1984.2 and 1989.2--1990.2.
Incidentally, these intervals cover the ejections and first detections
of C5 and C6,7.  In these time intervals, the $\nu_{\rm m}$ of the
core is rising, while the $S_{\rm m}$ is nearly constant, or even
decreasing slightly. According to the adopted model, it is impossible
to have $\varepsilon>0$ and $\rho <0$ simultaneously (see
Figure~\ref{fg:shockpar}).  This implies that $\nu_{\rm m}$ cannot
increase while $S_{\rm m}$ decreases (also ruling out fitting the
overall downward trend seen in the period 1982--90 in the core
spectrum). Figure~\ref{fg:shockpar} shows that $S_{\rm m}$ and
$\nu_{\rm m}$ can increase simultaneously only if $\rho\ge2.5$.  In
the two time intervals mentioned above, $\rho \sim 0$, so $\nu_{\rm
m}$ must decrease.  It is possible that the time sampling of the
spectral data is too poor to adequately represent the changes in the
core. Alternatively, the core may also represent a characteristic
region in relativistic flow. For instance, it can be the location at
which $\tau=1$ and the jet becomes optically thin. In this case, during
the time between two successive ejections of new jet features, the
spectral changes in the core are not necessarily consistent with those
of a shock.

\placefigure{fg:corevol}

\placetable{tb-SmNum}

The shock model is more successful in reproducing the decay stages
(1981--82, 1984--89) and the rise and fall in 1990-1993.
Table~\ref{tb-SmNum} gives the values of $\rho$, $b$ and $\varepsilon$
corresponding to the tracks A, B and C in Figure~\ref{fg:corevol}. In
the core, the jet axial coordinate $R$ has only a limited meaning,
since the core is believed to be stationary. One can possibly relate
$R$ to the time measured in the observer's frame, but the actual
proportionality is not obvious.  In view of that, the derived
$\delta(R)$ and $\nu_{\rm m}(R)$ dependencies can reflect the
magnitude of Doppler factor and turnover frequency variations only
indirectly. It is, however, symptomatic that the largest variations of
$\delta$ and $\nu_{\rm m}$ are required during the Compton--loss and
adiabatic--loss stages, whereas more moderate changes describe the
synchrotron--loss stages.

Assuming that the core is associated with the $\tau=1$ region of the
jet, the calculated $\rho$, $b$ and $\varepsilon$ may be reconciled with
the following scheme:

1990.2--91.8: a dense relativistic plasma condensation produced by an
outburst in the nucleus gradually becomes visible, as it enters the
$\tau=1$ region. If a shock is formed at this stage, the increased
plasma density, strong Compton losses and shock acceleration should
lead to increases in both $S_{\rm m}$ and $\nu_{\rm m}$. This is
reflected in $\rho>0$, $b>0$, $\varepsilon>0$.  If the initial density
of the plasma condensation is not large enough, the formation of a
shock may be delayed and $\nu_{\rm m}$ may rise without significant
changes in $S_{\rm m}$ (1982.4--84.2; 1989.3--90.2).

1991.8--92.5; 1984.2--85.8: the entire condensation is inside the core
region.  The plasma density remains constant (or decreases slightly),
and $b \sim 0$. Synchrotron losses dominate the emission, so that
$\nu_{\rm m}$ falls off ($\varepsilon < 0$), and $S_{\rm m}$ changes
weakly ($\rho \sim 0$). The most advanced parts of the condensation
begin to be seen as a separate feature receding from the core.

1992.5--93.8; 1985.8--89.2: the condensation leaves the core region;
the plasma density decreases rapidly. This can mimic an adiabatic
expansion, and cause both $\nu_{\rm m}$, and $S_{\rm m}$ to fall off
substantially. The steepest possible model slope $\rho=2.9$ (if
$b\rightarrow -\infty$) still falls short of representing the nearly
vertical downfall of $S_{\rm m}$ ($\rho \sim 10$) in 1985.8--89.2. If
the emission is still dominated by the synchrotron losses, it would
require the plasma number density, $N_0 \approx R^{-4} [N_0]_{\rm
adiabatic}$ to reproduce the data. It appears, therefore, that the
adiabatic expansion should be complemented by the recession of the
plasma condensation from the core.

\subsection{The components C4 and C5}

In the jet components, the changes of the $b$--parameter
required by the shock model can be compared with the values derived
from the component trajectories. The $S_{\rm m}$--$\nu_{\rm m}$ tracks
can also be tested against the observed component trajectories and
apparent speeds.

\subsubsection{$S_{\rm m}$--$\nu_{\rm m}$ evolution in C4 and C5}

The $S_{\rm m}$--$\nu_{\rm m}$ tracks of C4 and C5 are shown in
Figure~\ref{fg:turnevol}. The corresponding model parameters are
listed in Table~\ref{tb-SmNum}.  The data for C4 are too sparse. The
slope $S_{\rm m} \propto \nu_{\rm m}^{2.8}$ in 1983.4--88.2 gives
uncomfortably low values of $\varepsilon=-7.8$ and $b=-9.7$. The
derived $b$ is inconsistent with the component kinematics (Figure~\ref{fg:bmodel}). The last section of the component spectral track can
be approximated by a moderately accelerating shock in the adiabatic--loss stage.

The data for C5 are more complete, and the component's $S_{\rm
m}$--$\nu_{\rm m}$ track can be divided into three sections
corresponding to the main stages of shock development. Only the
1984.2--85.8 fraction of the track is problematic for fitting:
$\nu_{\rm m}$ is rising, while $S_{\rm m}$ is decreasing. In
connection with the spectral changes seen in the core in 1982--85, we
can speculate that a strong shock did not begin to form until about
1985. At later times, we can describe the spectral changes in C5 as
follows (see Table~\ref{tb-SmNum} for the corresponding model parameters):

\placefigure{fg:turnevol}

1985.8--88.2:~Compton--loss stage, with moderate acceleration
($b=2.0$). The component rest frame distance increases during this
period by $R_{88.2}/R_{85.8} \approx 1.1$. The kinematic value of the
$b$--parameter is $\approx1$ (adopting the $\gamma=\gamma_{\rm
min}(t)$ evolution of the component Lorentz factor; see
Figure~\ref{fg:bmodel}). If we take $b = 1$, then $S_{\rm m} \propto
\nu_{\rm m}^{14.2}$ which is too steep to match the data. 

1988.2--89.2:~synchrotron--loss stage, with constant Doppler factor
($b=0$).  Adoption of the kinematic $b(\gamma_{\rm min})\approx1$
would result in $S_{\rm m}\propto \nu_{\rm m}^{-8.2}$ which is
unrealistic. In one year, $\nu_{\rm m}$ changes from
$\approx7$\,GHz to $\approx3$\,GHz, which corresponds to
$R_{89.2}/R_{88.2} \approx 2.3$.

1989.2--92.5:~adiabatic--loss stage, with $b=-2.6$ ($R_{92.5}/R_{89.2}
\approx 1.6$). The $S_{\rm m}$--$\nu_{\rm m}$ slope in the data
appears to be steeper than the model value $\rho=1.8$, but the
increased errors allow for $\rho=0.9-2.6$. In Figure~\ref{fg:bmodel},
the $b(\gamma_{\rm min})$ of C5 decreases in 1989.2--92.5, but it
matches the model value only in 1992. If $\gamma_{\rm C5}$ is
constant, the corresponding $b$--parameter increases (as shown in
Figure~\ref{fg:bmodel} for $\gamma_{\rm C5}=7$; similar curves result from
other constant values of $\gamma_{\rm C5}$).

\subsubsection {Kinematic $b$--parameters of the jet components \label{sc:comkin}}

We derive the component trajectory and kinematic parameters using the
measured component position offsets ($x(t), y(t)$) from the core (see
ZCU95 for more details). Figure~\ref{fg:c5-track} shows the observed
offsets and the derived trajectory of C5. From the trajectory, the
component proper motion, $\mu(t) = [(dx/dt)^2 + (dy/dt)^2]^{1/2}$ and
apparent speed, $\beta_{\rm app}(t) = \mu(t) D_{\rm L} (1+z)^{-1}$ can
be calculated.  If the component Lorentz factor is known, the
corresponding three--dimensional trajectory can be
reconstructed. Several possible cases of the evolution of the Lorentz factor 
have been discussed in section \ref{sc:lum-kin}. In addition to those, we now
also consider the Doppler factors variations that can be deduced
from the observed spectral variations. For given $\gamma(t)$ or
$\delta(t)$, we can determine the corresponding $b$--parameter
variations, and compare them with the model predictions.  The
kinematic $b$--parameters must be measured locally, because of the
substantial and non--monotonic variations of the measured apparent
speeds. We use
\begin{equation}
\label{eq:bfactor}
b(t) = \frac{\log[\delta(t+dt)/\delta(t)]}{\log[R_{\rm j}(t+dt)/
R_{\rm j}(t)]}\, ,
\end{equation}
 where $R_{\rm j}$ is the rest frame traveled distance calculated 
from equation \ref{eq:Rj}. 

\placefigure{fg:c5-track}

\subsubsection{Peaked evolution of the turnover frequency}

We now compare the $b$--parameters derived from the observed tracks of
C4 and C5 with their counterparts obtained from modeling the spectral
changes in these components. The results of this comparison are
presented in Figure~\ref{fg:bmodel}.  The dot--dashed lines in
Figure~\ref{fg:bmodel} show the ranges of $b$--parameters for which a
peak in the time evolution of $\nu_{\rm m}$ is possible.  In both C4
and C5, the $b$--parameters corresponding to $\gamma_{\rm min}(t)$ lie
within the derived ranges. Therefore, for both C4 and C5, the
evolution of $\nu_{\rm m}$ which shows a peak, can in principle be
reconciled with the observed kinematic properties.
However, the choice of $\gamma_{\rm min}(t)$ is arbitrary, and is not
tied to the modeled spectral evolution.

Altogether, different phases of shock development can approximate the
spectral changes seen in C5. However, as seen in Figure~\ref{fg:bmodel}, the model $b$--parameters required by the adopted shock
evolution do not match well the kinematic values derived for both the 
$\gamma=const$ and $\gamma=\gamma_{\rm min}$ cases.  In the next
section, we will discuss a self--consistent scenario for C5, based on
both spectral and kinematic properties of the component.

\placefigure{fg:bmodel}

\subsubsection{Spectral and kinematic evolution of C5 \label{sc:c5-spkin}}

We now investigate how well a single set of model parameters can
describe the observed spectral and kinematic changes in C5.  If the
turnover frequency changes between two epochs ($t_1$, $t_2$) from
$\nu_1$ to $\nu_2$, then
\begin{equation}
\label{eq:R1}
R_1 = \Delta R_{\rm j} (\nu_1/\nu_2)^{1/\varepsilon} [ 1 - (\nu_1/\nu_2)^{1/\varepsilon} ]
\end{equation}
determines the location at which the component must be at $t_1$ in
order to satisfy the observed $S_{\rm m}(\nu_{\rm m})$ and the
$\nu_{\rm m}(R_{\rm j})$ required by the model.  We consider two cases
in which the initial Doppler factors are: $\delta_{01} = 1.05$ and
$\delta_{02} = 2.8$. For these cases, we determine the $R_1$ at each
stage of the spectral evolution in C5 (Table~\ref{tb-c5stages}). The
derived $R_1$ are much smaller for the synchrotron--loss stage
(1988.2-89.2), compared to the values for the other two stages. This
indicates that the spectral changes occurring in C5 during this period
are too rapid to satisfy both the shock model and the observed
kinematics. No satisfactory fit can be achieved by varying the model
$S_{\rm m}(\nu_{\rm m})$ slopes and durations of each of the stages of
the evolution of the shock.  Assuming that spectral points with $5<\nu_{\rm
m}<15$\,GHz are the most accurate, we reproduce a self-consistent
$S_{\rm m}$--$\nu_{\rm m}$ track (the dot--dashed line in
Figure~\ref{fg:turnevol}) which satisfies the observed trajectory of
C5. It is clear, from Figure~\ref{fg:turnevol}, that the modeled track
cannot accommodate the rapid spectral changes in 1988-92.  Variations in
the initial Doppler factor $\delta_0$ do not change the model track
significantly. The shape of the track is chiefly determined by the
rise in 1985-88. In that period, the only plausible deviation from the
modeled shape is to make the rising slope shallower, and to attempt
extending the flat section of the track (corresponding to the
synchrotron--loss stage) by varying the $b$--parameter. The rising slope
in 1985--88 can be made shallower only by forcing $\rho<5.1$.  As a
result, the corresponding values of $b$ grow significantly larger, and
the length of the rising section of the track grows dramatically. This
precludes finding any satisfactory fit for the synchrotron and adiabatic
stages of the shock, leaving no room for a successful shock--based scenario
capable of explaining both the spectral and the kinematic properties of C5.

The difficulties in explaining the observed spectral and kinematic
properties of C5 may be caused by several different factors. One
possible explanation is that C4 and C5 could have been produced by
the same outburst in 3C\,345. The estimated ejection epochs of C4 and
C5 differ only by about 0.5 year, yet the emission and kinematic
properties of the two components are significantly different (L96). C4
is much faster and brighter, compared to C5. It is therefore possible
that C4 and C5 represent the forward and reverse shocks propagating in
the jet.\footnote{The first anonymous referee of the manuscript
suggested that C5 could also be traveling through the plasma that has
been shocked by C4. This could weaken the emission of C5, and alter
its spectral evolution.} There is also observational evidence for a
possible transition occurring in the jet of 3C\,345 at about
1.2--1.5\,mas distance from the core \cite{lz96}. At this distance,
the component proper motions, polarization properties and flux density
behavior change. For the jet, this may indicate a change in the dominating
emission mechanism (for instance, if at this distance strong shocks
have already dissipated, and the jet emission properties become
determined largely by interactions between the jet and the ambient
medium). C5 has passed the 1.5\,mas point in 1988.1; incidentally,
this marks the time when applying the shock model to the spectral
evolution of C5 becomes difficult.

\placetable{tb-c5stages}

\section{The VLBI core as an inhomogeneous relativistic jet \label{sc:konigl}}

As follows from section \ref{sc:core}, relativistic shocks have
noticeable difficulties reproducing the observed spectral changes in
the VLBI core of 3C\,345 within a single, self--consistent
scheme. Variations of the particle number density---a free parameter in
the shock model---appear to be necessary to fit the observed values.
With this requirement in mind, we now apply, to the VLBI core, a
modification of the quasi--steady, inhomogeneous relativistic jet
model proposed by K\"onigl (1981) to describe the emission from
ultra--compact jets.  In its original formulation, the model is
stationary (i.e., it does not make specific predictions about the time
variability of the emission), and because of this, it has been
previously applied for fitting the core spectrum in 3C\,345 only at
single epochs: 1982.0 (ZCU95) and 1992.7 (Unwin et al. 1997).

In order to permit self--consistent fits to multi--epoch data, we need
to make several assumptions about those model parameters which are
unconstrained by the spectral data.  We assume that a relativistic
plasma with spectral index $\alpha_0=-0.5$ is continuously injected
into the jet with bulk Lorentz factor $\gamma_{\rm j}$.  The jet is
unresolved (which also follows from the measured sizes of the core, as
reported in ZCU95 and L96). The jet geometry is approximated by a cone
with an opening angle $\phi_{\rm j}\approx const$. The axis of the
cone forms an angle $\theta_{\rm j}\approx const$ with the line of
sight. The jet plasma is characterized by a tangled magnetic field, $B
= B_1 R^{-a}$, and particle density, $N = N_1 R^{-n}$ (here $B_1$ and
$N_1$ are measured at $R=1$\,pc). The number of particles is
conserved, and the particle energy density in the jet, $u_{\rm p}$, is
in approximate equipartition with the magnetic field energy density,
$u_{\rm B}$, so that $u_{\rm p} = q_{\rm e} u_{\rm B}$ (with $q_{\rm
e} \sim 1$, but possibly varying from one epoch to another).  This
implies that $a=1$, $n=2$, and $N_1 = q_{\rm e} B_1^2 (8\pi m_{\rm e}
c^2)^{-1}$ (where $m_{\rm e}$ is the electron mass, and $c$ is the
speed of light). Finally, we must take into account the effect of
changes in the optical depth, from one epoch to another, expressed by the
distance, $r_{\rm m} \propto
\delta^{-1/6} N_1^{1/6} B_1^{11/6}$, at which the jet becomes optically
thin for the synchrotron radiation. With this additional requirement (and
for the choice of parameters described above), the observed variations
of the turnover point of the synchrotron spectrum of the VLBI core should
comply with the following relations:
\begin{equation}
\label{eq:konSm}
S_{\rm m} = C_{\rm S} N_1^{0.42} B_1^{4.08} \delta^{5.08}\,,
\end{equation}
\begin{equation}
\label{eq:konVm}
\nu_{\rm m} = C_{\nu} N_1^{0.17} B_1^{-0.17} \delta^{0.83}\,.
\end{equation}
To simplify further discussion, we will drop $C_{\rm S}$ and $C_{\nu}$
(both of them vary weakly with respect to $\gamma_{\rm j}$, and are
constant with respect to our choice of other model parameters). We
then consider only fractional changes of $N_1$, $B_1$, and $\delta$,
by taking their respective values at the first spectral epoch ($t_0 =
1981.5$) as unity.  One has to make an assumption about the changes of
one of these three quantities, in order to reconstruct the evolution
of the other two.  We take $\delta = const$ as the most plausible
assumption, relying again on the similarity between the measured
apparent jet speeds in the immediate vicinity of the core (ZCU95,
L96).  This assumption makes it possible to determine the changes of $N_1$
required to reproduce the observed spectral variations. These changes
are plotted with open circles in Figure~\ref{fg:konigl1}. The relative
peak--to--peak variations of $N_1$ do not exceed 500. The
corresponding magnetic field varies within 20\%, making the proposed
scheme rather attractive. We therefore conclude that moderate
variations of the particle density in an unresolved, inhomogeneous jet
with constant Doppler factor can explain the observed spectral
variations in the VLBI core of 3C\,345.

\placefigure{fg:konigl1}

\subsection{Flare model for the particle density variations}

We will attempt now to find out whether the determined changes of
$N_1$ can be accommodated within a simple and self--consistent
description of density variations in the jet. To represent the changes
of $N_1$, we use flare--like events described by an exponential rise
and decay of the density, so that $N(t) = N_{\rm rel} \exp(t_{\tau})$,
where $t_{\tau} = |t - t_{\rm flare}| \tau_{\rm flare}^{-1}$. Here
$t_{\rm flare}$ and $\tau_{\rm flare}$ refer to the epoch and duration
of the flare, and $N_{\rm rel} = N_{\rm flare}/N_{\rm quiescent}$ is
the relative density increase during a flare.  We take $N_{\rm
quiescent}=1$, and find that the changes of $N_1$ determined in the
previous section can be represented by a sequence of 5 flares
occurring roughly every 3.5--4 years. The parameters of the flares are
listed in the ``spectrum'' section of Table~\ref{tb:flares}.  The
density variations resulting from these flares are represented by the
dotted line in Figure~\ref{fg:konigl1}. Figure~\ref{fg:konigl1} also
shows the variations of the Doppler factor (open squares) and magnetic
field (filled triangles) that correspond to the sequence of the flares
obtained.  The epochs of the flares and their relative strengths are
shown in Figure~\ref{fg:konigl1} by stars and segments. We are not
able to find a satisfactory fit to the density at the last spectral
epoch (1993.8), without introducing a flare in 1995. Since the flares
in 1981 and 1995 are not constrained by the spectral data, we use
dashed--line trapeziums to illustrate the allowed ranges of epochs and
strengths of these flares.  The rather large values of $N_{\rm rel}$
suggest that the quiescent particle density in the jet is likely to be
small: the jet material should then be supplied entirely by the
flares. In this case, one can expect a strong correlation between the
occurrence of the flares and variability of the total radio emission.
To illustrate this, we plot also the 22\,GHz light curve in
Figure~\ref{fg:konigl1} (in order to better place the light curve
within the plot, we scale the fluxes down by a factor of 100). The
total flux evolution agrees well with the flare model: the much weaker
flare at 1987.8 has resulted in the dramatic decline of the total
emission (around 1990, the radio emission from 3C\,345 was at its
all--time low). Conversely, the subsequent, stronger flares in 1992
and 1995 are noticeably reflected by the increased total flux density.

With regard to the scheme presented above, we would like to stress
that the suggested sequence of the flares is not strictly a fit to the
data. Sparsity of the spectral epochs does not warrant proper
fitting. In addition, our flare model is based on the assumption of
$\delta=const$ which may (in general) be relaxed, possibly altering
the derived flare parameters.  However, we find it rather remarkable
and attractive that varying only a single parameter ($N_1$) of the
model can be sufficient for explaining the entire observed spectral
evolution in the VLBI core, and that the variations of $N_1$ can be
represented by a rather regular sequence of flare--like events
occurring in the nucleus of 3C\,345.

\placetable{tb:flares}

\subsection{Applying the flare model to the light curves}

We now attempt to apply the flare model to reproduce the emission
variations at 22\,GHz. The total emission is not well suited for this
purpose, because it contains, even at 22\,GHz, significant (20--50\%)
contributions from the jet components.  We therefore choose to fit the
flare model to the flux density variations of the VLBI core.
The available flux density data for the VLBI core
cover the time period from 1981 until 1997, continuing for 3 years
after the last spectral epoch (1993.8), and sampling well the flare in
1995.

To describe the flux density changes during a flare, we consider the
synchrotron emission from a stationary adiabatic flow, confined by the
pressure, $p\propto R^{-4\epsilon}$, of the external medium
(Georganopoulos \& Marscher 1996). The emission can then be described by
equation~\ref{eq:synspec}, with $\nu_1(t)$ incorporating the adiabatic
and synchrotron energy losses in the plasma:
\begin{equation}
\label{eq:V1}
\nu_1 (t) = \nu_{\rm 1,0} \left[\left( 1 + \frac{\beta_{\rm j}\delta_{\rm j}}{R_{\rm core}}
(t - t_{\rm flare}) \right)^{(2\alpha -3)\epsilon} N_{\rm rel} \exp(t_{\tau})\right]^{2/(5-2\alpha)}\,.
\end{equation}
We assume $\nu_{1,0}=10$\,GHz and $\epsilon=1/3$. The kinematic
conditions are described by $R_{\rm core} \approx3.5$\,pc (Lobanov
1998a) and $\delta_{\rm j} \approx 6$ (from $\gamma_{\rm j}\approx 3.5$ and
$\theta_{\rm j}\approx 5^{\circ}$ typically measured in the vicinity
of the VLBI core). We take the previously obtained flare sequence as a
starting model, and adjust the flare parameters so as to fit the core
flux densities.  The resulting fit gives a remarkably good
approximation of the observed flux density changes in the core. We plot the
obtained fit in Figure~\ref{fg:konigl2}, together with the variations
of the total and VLBI core flux densities. The parameters of the
flares are given in the ``light curve'' section of
Table~\ref{tb:flares}.

The discrepancies between the epochs and durations of the ``spectral''
and ``light curve'' flares are small, with exception of the weak flare
in 1986--87 for which fitting is difficult in both cases.  The
``spectral'' flare in 1992.5 is not constrained by the data, which can
explain why its density increase differs from that found in its
counterpart ``light curve'' flare.  The densities inferred from the
``light curve'' flares give a better representation of the density
changes in the jet, since these flares are derived from the measured
values of the flux density (in contrast to the ``spectral'' flares
which describe the modeled relative changes of the particle density). The
density increase in the ``light curve'' flares is up to 10 times
larger than in their ``spectral'' flare counterparts, further
enhancing the argument that flares supply the major fraction of the
material carried by the jet.

Noticing the remarkable overall regularity of the flare events in
3C\,345, we can venture to predict that the next flare (already
visible in the 22\,GHz light curve in Figure~\ref{fg:konigl2}) should
peak at around mid--1999, with the total flux density reaching at
least 16\,Jy.  It is conceivable that the flare sequence seen in
3C\,345 results from physical settings similar to those uncovered in
OJ\,287 (Valtonen \& Lehto 1997), where the observed
quasi--periodical variations of radio emission are likely to be caused by
the orbital motion in a binary black hole system.  This argument,
however, lies outside the scope of this paper, and will be discussed
elsewhere.

We limit ourselves here to remarking again on the simplicity and 
success of the obtained flare model which, for both spectral and flux
density data, provides a satisfactory description relying only on
variations of the particle density in the jet. The similarity of
``spectral'' and ``flux density'' flares also provides an additional
argument for the fidelity of the spectral information obtained from
the VLBI data.

\placefigure{fg:konigl2}

\section{Summary}

We have used the available multi--frequency VLBI observations to
determine the basic properties of the synchrotron spectra of the
enhanced emission regions in the parsec--scale jet of 3C345. The
spectral information has been combined with the kinematic data, in
order to derive suitable models for the observed evolution of the
radio emission from the VLBI core and the moving components in the jet.

1.~We have demonstrated that the compact jet emission can be
adequately modeled by Gaussian components, used for fitting the
interferometric visibility data obtained from VLBI observations.
The models obtained can be combined into multi--frequency datasets,
providing the necessary basis for studying the synchrotron spectra
in the parsec--scale jet of 3C\,345. 

2.~The turnover frequency and the flux density can be recovered from
multi--frequency datasets, using the measured curvature of the
spectrum within the range of the observing frequencies, $\nu_{\rm
l}$--$\nu_{\rm h}$. We have studied the limitations of the method and
shown that it gives reliable results for turnover frequencies between
$\sim0.3\nu_{\rm l}$ and $\sim 1.3\nu_{\rm h}$, for data with flux
density errors not exceeding 10\%.  Using this method, we have fitted
the synchrotron spectra and estimated the turnover frequencies and the flux
densities (or respective upper limits), for the VLBI core and the jet
components in 3C\,345.

3.~From the spectra obtained, we estimated the 4--25\,GHz luminosities
of the core and the jet components.  The mean luminosity of the core
is $\approx7\cdot 10^{42}\,h^{-2}$erg\,s$^{-1}$. The total luminosity
of the parsec--scale jet in 3C\,345 is $\approx3\cdot
10^{43}\,h^{-2}$erg\,s$^{-1}$ which constitutes about 1\% of the
observed luminosity of the source in the radio to infrared regime.
The observed luminosity changes indicate that the core and the jet
components have comparable energy loss rates, and suggest that the jet
components are accelerating in the rest frame of the jet.

4.~The spectral evolution of the VLBI core differs from the changes
seen in the spectra of the jet components.  The spectral changes
observed in the core can be partially reproduced by a relativistic
shock traveling through the region where the jet turns optically
thin. A better description is provided by the framework of the
inhomogeneous jet model which is capable of explaining the entire
spectral evolution of the core by a sequence of flare--like events
characterized by an exponential rise and decay of the particle density
of the jet plasma.  The same model is sufficient for explaining the
observed flux density variations in the core of 3C\,345. The model
requires changes only one model parameter---the particle density in
the jet---in order to reproduce the measured spectral and flux density
data. The particle density variations obtained suggest that flares are
likely to be supplying most of the material carried by the
parsec--scale jet. The flares occur approximately once every 3.5--4
years, suggesting that a quasi--periodic process in the nucleus of
3C\,345 may be driving the flaring activity. On the basis of the flare
sequence determined, we can expect that the current flare, which began
at the end of 1997, should peak in mid--1999, with the total flux
density reaching at least 16\,Jy.

5.~There is a peak in the evolution of the turnover frequency of C5 (and
possibly, C4). The peak in C5 can be reconciled with a moderately
accelerating shock. The data are too poor to make any conclusion about
C4. The kinematic properties of both C4 and C5 allow for the
variations of the component Doppler factors required to produce a peak
in the turnover frequency evolution.  The shock--in--jet model has
difficulties with reproducing the overall spectral and kinematic
changes observed in C5. The spectral evolution is too rapid
(especially at the later epochs) to be accommodated by the model,
within the kinematic constraints determined by the component's
observed trajectory. It is conceivable that the later epochs are
marked by a rapid dissipation of the shock, although it remains
difficult to single out the most likely cause for such an
effect.

\section{Acknowledgments}

We would like to acknowledge P.E.~Hardee, K.I.~Kellermann, T.P.~Krichbaum,
A.P.~Marscher and J.~Roland for fruitful discussions during the course
of this work.  We thank I.I.K.~Pauliny--Toth for many helpful comments
on the manuscript. We also wish to thank the anonymous referees for
numerous useful suggestions. We would like to acknowledge
J.A.~Biretta, M.H.~Cohen, K.J.~Lepp\"anen, S.C.~Unwin, and A.E.~Wehrle
for their valuable contributions to the long--term VLBI monitoring of
3C\,345. JAZ acknowledges partial support by the Alexander von
Humboldt Foundation's Forschungspreis.  The National Radio Astronomy
Observatory is a facility of the National Science Foundation operated
under cooperative agreement by Associated Universities, Inc.

\newpage

\figcaption{ The top image: VLBI image of 3C345 at 3.6cm (Lobanov 1996). 
The contour levels are $-$0.1, 0.1, 0.2, 0.3, 0.5, 1, 2, 3, 5, 10, 25,
50, 75, and 90 \% of the peak, 4.952\,Jy. The bottom image: Gaussian
model fit representation of the VLBI image, obtained by fitting 8
elliptical Gaussian components to the visibility data and optimizing
the fit to the amplitudes and closure phases. Labeled are the VLBI
stationary core D, and the components C7, C6, C5, C4, C3 and C2
embedded in the jet. The weakest and most extended jet feature, C1,
lies outside the plotting range, at a distance of $\sim
20$\,mas.\label{fg:36nov93map}}

\figcaption{Example of fits to VLBI visibility data. 
The visivility amplitudes on two baselines (top) and closure phases on two
triangles (bottom) are taken from the VLBI dataset used for producing the
images in Figure~1. In the left panels, CLEAN component model is fitted
to the visibilities; the right panels show the fit obtained by modeling
the source structure with 8 elliptical Gaussian components. Consistency
of the fits by the CLEAN and Gaussian model warrants the use
of Gaussian model fits for describing the mas--scale structures in 3C\,345.
\label{fg:visfit}}

\figcaption{Theoretical curvature, $\kappa$, of the functional form
(1) describing the homogeneous synchrotron spectrum with spectral
index $\alpha$. The $\xi$ axis is the ratio of the frequency at
which the curvature is calculated to the turnover frequency. Different
symbols mark the smallest measurable curvature, $\kappa_{\rm min}$,
for a spectral dataset with the given fractional flux density
error. The meaningful corrections can be achieved for values of
$\kappa \le \kappa_{\rm min}$ \label{fg:kappa}}

\figcaption{Spectral fits to the C5 data. Dotted lines are third order
polynomial fits used for measuring the local curvature. Solid lines
are the derived synchrotron spectra. The curvature is measured at the
lowest observed frequency, unless the turnover is seen in the data
(epochs 1987.3, 1988.2). In the latter case, the curvature is measured
at the turnover frequency derived from the corresponding polynomial
fit. When only two spectral points are available (epochs 1991.7, 1993.8), the
fitted values represent the upper limits on the turnover
frequency. \label{fg:c5fits}}

\figcaption{Evolution of the turnover frequency $\nu_{\rm m}$ in the core
and the jet components C4 and C5. For each feature, arrows mark the
upper limits.
\label{fg:newevol}}

\figcaption{Evolution of the turnover flux density $S_{\rm m}$ in the core
and the jet components C4 and C5. For each feature, arrows mark the
lower limits.
\label{fg:turnflux}}

\figcaption{Evolution of the luminosity in the VLBI core and the 
jet components. The shown luminosities are the upper limits derived
from the measured integrated fluxes in the range of 4--25\,GHz,
$S_{\rm int}$, and estimated minimum Lorentz factors, $\gamma_{\rm
min}$, listed in Table~2.
\label{fg:evolbeta}}

\figcaption{Luminosities and kinematics of the jet components: 
{\bf a)}~luminosities and apparent traveled distances; the dot--dashed
line is a linear fit to C4 data; the solid line represents a linear
fit for C2, C3, and C5 data combined.  {\bf b)}~luminosities and rest
frame traveled distances; the components move at constant speeds,
$\gamma_{\rm C2,3} = 20$, $\gamma_{\rm C5} = 7$, $\gamma_{\rm C4} =
15$.  {\bf c)}~the same as b), but each component moves at its
$\gamma_{\rm min}(t)$.  {\bf d)}~variations of $\gamma_{\rm min}(t)$
in the jet components in the rest frame of the jet.
\label{fg:loom}}

\figcaption{Parameter space of the shock--in--jet model. The power
indices $\rho$ ($S_{\rm m} \propto \nu_{\rm m}^{\rho}$) and $\varepsilon$ 
($\nu_{\rm m} \propto R^{\varepsilon}$) are shown for different values of
$b$ ($\delta \propto R^{b}$), for the three basic stages of the evolution of
the shock. The dependencies shown are calculated for $a=1$ and $s=2.4$;
the respective dependencies for $1<a<2$ and $1<s<3$ do not differ
significantly from the plotted curves.\label{fg:shockpar}}

\figcaption{$S_{\rm m}$--$\nu_{\rm m}$ changes in the core. The lines
and letters indicate different stages of the evolution of the shock,
as described in section~4.1 and Table~5.
\label{fg:corevol}}

\figcaption{$S_{\rm m}$--$\nu_{\rm m}$ changes in C4 and C5. For each 
component, a dotted line shows possible tracks consistent
with different stages of the evolution of the shock (the corresponding model
parameters are given in Table~5).  The dot--dashed line shows how the
original fit for C5 must be changed to satisfy the observed
trajectory of the component (plotted in Figure~13).\label{fg:turnevol}}

\figcaption{The observed trajectory of C5 in the plane of the sky. The 
dotted line represents the combined polynomial fits to the component's
$x$ and $y$ offsets from the core. Open circles mark the locations on
the trajectory which are equally spaced in time at an interval of 1
year.\label{fg:c5-track}}

\figcaption{The kinematic and shock--model $b$--parameters for C4 and C5. 
Different symbols show the $b$--parameters determined for the
$\gamma=const$ (triangles) and $\gamma=\gamma_{\rm min}$ (circles)
kinematic solutions. Solid lines represent the changes of the
$b$--parameters of the components, as described in Table~5.
Dot--dashed lines indicate the ranges of $b$--parameters for which a
the turnover frequency can have a peak in time during the
Compton--loss stage of the evolution of the shock.
\label{fg:bmodel}}

\figcaption{Relative changes of the Doppler factor and magnetic field in 
 the VLBI core, obtained by applying K\"onigl jet model to the
 measured $S_{\rm m}$ and $\nu_{\rm m}$. The variations of all
 quantities are normalized to their respective values at the first
 epoch, $t_0=1981.5$. Open circles denote the changes of the particle
 density required for maintaining a constant Doppler factor. The
 dotted line shows the variations of the particle density, as
 represented by 5 exponential flares (the parameters of the flares are
 listed in the section ``spectrum'' of Table~7). The resulting
 variations of the Doppler factor (open squares) and magnetic field
 (filled triangles) are also shown, with lines representing linear
 fits to the respective quantity. The epochs of the flares are
 indicated in the bottom by the stars. For each flare, a segment
 indicates the relative increase of the particle density during the
 flare. The first and the last flares are not constrained by the data;
 the trapezia indicate the acceptable ranges of the locations and the
 amplitudes of these flares. The total 22\,GHz flux density light
 curve (Ter\"asranta 1998 and priv.comm.) is plotted for
 comparison. The light curve is scaled down by a factor of 100.
\label{fg:konigl1}}

\figcaption{Application of the flare model to the observed variations
 of the 22\,GHz flux density of the VLBI core in 3C\,345. Filled
triangles are the measured flux densities of the core. The solid line
is the fit by the flare model.  The parameters of the flares are
listed in the section ``light curve'' of Table~7. If the current flare
is similar to the flare in 1992, the dashed line should represent the
expected evolution ofthe flux density in the core.  Open circles show
the changes in the total emission at 22\,GHz (Ter\"asranta 1998 and
priv.comm.).  The flux densities of the core for the period
1994--1997 are taken from Lepp\"anen (1995) and Ros et al. (1999).
\label{fg:konigl2}}

\newpage


\begin{table}
\caption{Multi frequency data sets.\label{tb-dat}}
\footnotesize
\begin{center}
\medskip
\begin{tabular}{cccccc} \hline\hline
(1) & (2) & (3) & (4) & (5) & (6) \\
$\tau_\nu$ & $t_{obs}$ & $\nu$ & $S_{VLBI}$ & $S_{tot}$ & Ref. \\ \hline\hline 
 & 1979.25 &  4.99 & $ 7.50\pm 0.41$ & ... & 1 \\
1979.3 & 1979.44 & 10.70 & $ 8.53\pm 1.28$ & $ 8.59\pm 0.14$ & 1,9 \\ 
\hline
 & 1980.52 & 10.70 & $10.48\pm 1.93$ & $10.13\pm 0.10$ & 1,9 \\
1980.6 & 1980.73 &  4.99 & $ 7.40\pm 0.15$ & $ 8.02\pm 0.05$ & 1,9 \\ 
\hline
 & 1981.09 & 10.70 & $12.83\pm 1.24$ & $13.01\pm 0.60$ & 1,9 \\
 & 1981.25 & 22.30 & $13.87\pm 0.30$ & $14.92\pm 0.77$ & 1,10 \\
 & 1981.63 &  4.99 & $ 9.89\pm 0.17$ & $10.08\pm 0.05$ & 1,9 \\ 
1981.5 & 1981.68 & 2.30 & $6.27\pm0.67$ & ... & 1 \\ 
\hline
 & 1982.09 & 10.70 & $15.19\pm 0.53$ & $15.34\pm 0.32$ & 1,9 \\
 & 1982.38 & 89.20 & $12.00\pm 6.40$ & $13.47\pm 8.62$ & 1,10 \\
1982.4 & 1982.56 &  4.99 & $10.67\pm 0.30$ & $11.35\pm 0.07$ & 1,9 \\
\hline
 & 1983.09 & 22.30 & $13.20\pm 0.90$ & $15.01\pm 0.28$ & 1,10 \\
 & 1983.10 & 10.70 & $14.66\pm 1.06$ & $14.67\pm 0.26$ & 1,9 \\
1983.4 & 1983.57 &  4.99 & $ 9.52\pm 0.87$ & $11.44\pm 0.15$ & 1,9 \\
\hline
 & 1984.09 & 22.30 & $13.20\pm 0.64$ & $13.87\pm 0.23$ & 1,10 \\
 & 1984.11 & 10.70 & $13.74\pm 0.84$ & $14.12\pm 0.31$ & 1,9 \\
1984.2 & 1984.25 &  4.99 & $ 7.43\pm 0.38$ & $12.80\pm 0.12$ & 2,9 \\
\hline
 & 1985.75 & 22.30 & $ 6.36\pm 0.18$ & $10.90\pm 0.28$ & 2,10 \\
 & 1985.77 &  4.99 & $ 9.25\pm 0.24$ & $11.52\pm 0.06$ & 2,9 \\
1985.8 & 1985.93 & 10.70 & $ 9.88\pm 0.72$ & $11.44\pm 0.79$ & 2,9 \\
\hline
\end{tabular}
\medskip
\end{center}
\end{table}

\addtocounter{table}{-1}

\begin{table}
\caption{(continued)}
\footnotesize
\begin{center}
\medskip
\begin{tabular}{cccccc} \hline\hline
(1) & (2) & (3) & (4) & (5) & (6)  \\
$\tau_\nu$ & $t_{obs}$ & $\nu$ & $S_{VLBI}$ & $S_{tot}$ & Ref. \\ \hline\hline
 & 1986.90 &  4.99 & $10.01\pm 0.82$ & $10.56\pm 0.08$ & 2,9 \\
 & 1987.15 & 10.70 & $ 9.47\pm 0.16$ & $11.03\pm 0.48$ & 2,9 \\
1987.3 & 1987.42 & 22.30 & $ 3.93\pm 0.10$ & $ 9.32\pm 0.21$ & 2,10 \\
\hline
 & 1988.16 & 22.30 & $ 5.07\pm 0.09$ & $ 7.73\pm 0.18$ & 2,10 \\
 & 1988.17 & 10.70 & $ 8.70\pm 0.78$ & $ 8.86\pm 0.52$ & 2,9 \\
 & 1988.18 &  4.99 & $ 7.84\pm 0.12$ & $ 9.52\pm 0.09$ & 2,9 \\
1988.2 & 1988.21 & 100.00 & $3.20\pm 0.85$ & ... & 3 \\
\hline
 & 1989.22 & 100.00 & $3.20\pm 0.52$ & ... & 3 \\
 & 1989.25 & 22.30 & $ 5.98\pm 0.30$ & $ 5.89\pm 0.21$ & 7,9 \\
 & 1989.26 & 10.70 & $ 7.53\pm 0.56$ & $ 7.01\pm 0.33$ & 7,8 \\
1989.2 & 1989.28 &  4.99 & $ 7.85\pm 0.45$ & $ 8.06\pm 0.06$ & 4,8 \\
\hline
 & 1990.16 & 10.70 & $ 3.14\pm 0.16$ & $ 5.81\pm 0.03$ & 6,8 \\
 & 1990.18 &  4.99 & $ 6.50\pm 0.35$ & $ 6.56\pm 0.06$ & 4,8 \\
 & 1990.42 & 22.30 & $ 3.01\pm 0.22$ & $ 5.92\pm 0.20$ & 6,9 \\
1990.3 & 1990.49 & 43.20 & $ 7.88\pm 0.76$ & $ 5.04\pm 0.45$ & 5,8 \\
\hline
 & 1991.68 & 43.20 & $22.70\pm 4.54$ &  ... & 5 \\
 & 1991.71 &  4.99 & $ 4.97\pm 0.82$ & $ 5.24\pm 0.05$ & 4,8 \\
1991.7 & 1991.86 & 22.30 & $12.10\pm 0.72$ & $12.36\pm 0.28$ & 7,9 \\
\hline
 & 1992.43 &  4.99 & $ 6.61\pm 0.74$ & $ 6.43\pm 0.04$ & 7,8 \\
 & 1992.45 & 22.30 & $13.23\pm 2.04$ & $11.97\pm 0.43$ & 7,9 \\
1992.5 & 1992.71 &  8.40 & $10.59\pm 4.08$ & $10.35\pm 0.15$ & 7,8 \\
\hline
\end{tabular}
\medskip
\end{center}
\end{table}

\addtocounter{table}{-1}

\begin{table}
\caption{(continued)}
\footnotesize
\begin{center}
\medskip
\begin{tabular}{cccccc} \hline\hline
(1) & (2) & (3) & (4) & (5) & (6)  \\
$\tau_\nu$ & $t_{obs}$ & $\nu$ & $S_{VLBI}$ & $S_{tot}$ & Ref. \\ \hline\hline
 & 1993.70 &  4.99 & $ 6.42\pm 0.46$ & $ 8.15\pm 0.06$ & 7,8 \\
 & 1993.72 & 22.30 & $ 6.14\pm 0.16$ & $ 8.53\pm 0.31$ & 7,9 \\
1993.8 & 1993.88 &  8.40 & $ 8.48\pm 0.20$ & $ 9.99\pm 0.19$ & 7,8 \\
\hline
\end{tabular}
\end{center}
\tablecomments{1 -- multi frequency data set epoch; 2 -- observation epoch; 3
-- frequency, GHz ; 4 -- VLBI flux density , Jy; 5 -- total flux
density, Jy; 6 -- references for the VLBI and total flux density
measurements (for the 10.7 and 43.2\,GHz VLBI observations, the table 
gives the flux densities at 8 and 37\,GHz respectively).}

\tablerefs{ 1 -- Biretta et al.\, 1986; 2 --
Zensus et al.\, 1995; 3 -- B{\aa}{\aa}th et al.\, 1992;
4 -- Unwin \& Wehrle 1992;
5 -- Krichbaum et al.\, 1993; 6 -- Unwin et al. 1994;
7 -- Lobanov 1996; 8 -- Aller et al.\, 1996; 9 -- Ter\"asranta et al.\, 1998.}
\end{table}

\begin{table}
\caption{Spectral fits. \label{tb-fits}}
\footnotesize
\begin{center}
\medskip
\begin{tabular}{ccccrrrrrr} \hline\hline
(1) & (2) & (3) & (4) & (5) & (6) & (7) & (8) & (9) & (10)\\
Name      & $\tau_\nu$  & N$_\nu$  & $\sigma_{\rm S,N}$  & $S_{\rm int}$ & $S_m$ &  $\nu_m$ & $\alpha_{\rm fit}$ & $\gamma_{\rm min}$ & L$^{4-25\,GHz}_{\rm max}\,h^{-2}$  \\ \hline\hline
D & 1979.3 & 2 &  6.1  & $81.8\pm 3.7$    & $(4.7)$       & $(10.5)$      & -0.1 &(3.5)& $8.0\cdot 10^{42}$\\ 
  & 1980.6 & 2 & 15.9  & $82.6\pm 3.2$    & $(4.6)$       & $(8.5)$       & -0.2 &(3.5)& $8.0\cdot 10^{42}$\\
  & 1981.5 & 4 &  8.5  & $164.9 \pm 8.9$  & $10.1\pm 5.8$ & $14.9\pm 1.2$ & -0.1 &(3.5)& $1.2\cdot 10^{43}$\\
  & 1982.4 & 3 &  9.0  & $135.3 \pm 7.5$  & $7.4\pm 1.4$  & $ 6.7\pm 0.6$ & -0.3 &(3.5)& $1.3\cdot 10^{43}$\\
  & 1983.4 & 3 &  6.8  & $113.1 \pm 5.7$  & $5.6\pm 1.5$  & $10.1\pm 1.0$ & -0.2 &(3.5)& $9.2\cdot 10^{42}$\\
  & 1984.2 & 3 &  4.1  & $117.6 \pm 6.0$  & $6.8\pm 3.1$  & $22.5\pm 1.1$ & -0.1 &(3.5)& $7.9\cdot 10^{42}$\\
  & 1985.8 & 3 &  7.7  & $ 71.3 \pm 2.4$  & $3.7\pm 0.7$  & $12.1\pm 0.6$ & -0.2 &(3.5)& $5.8\cdot 10^{42}$\\
  & 1987.3 & 3 &  13.7 & $ 44.8 \pm 1.3$  & $2.3\pm 0.6$  & $12.1\pm 1.6$ & -0.1 &(3.5)& $3.7\cdot 10^{42}$\\
  & 1988.2 & 4 &  9.5  & $ 33.8 \pm 1.0$  & $1.9\pm 0.6$  & $13.2\pm 0.9$ & -0.1 &(3.5)& $2.9\cdot 10^{42}$\\
  & 1989.2 & 4 &  6.4  & $ 29.7 \pm 1.2$  & $1.4\pm 0.3$  & $10.3\pm 4.3$ & -0.2 &(3.5)& $2.5\cdot 10^{42}$\\
  & 1990.3 & 4 &  5.4  & $ 31.9 \pm 1.1$  & $1.6\pm 0.6$  & $15.6\pm 6.4$ & -0.1 &(3.5)& $2.5\cdot 10^{42}$\\
  & 1991.7 & 3 &  6.9  & $106.0 \pm 4.1$  & $7.3\pm 3.5$  & $32.5\pm 6.8$ & -0.1 &(3.5)& $6.3\cdot 10^{42}$\\
  & 1992.5 & 3 &  9.9  & $195.6 \pm 6.7$  & $11.9\pm 4.3$ & $24.0\pm 1.2$ & -0.1 &(3.5)& $1.2\cdot 10^{43}$\\
  & 1993.8 & 3 &  2.4  & $ 66.4 \pm 3.0$  & $4.0\pm 1.6$  & $20.0\pm 1.8$ & -0.1 &(3.5)& $4.4\cdot 10^{42}$\\ \hline
\end{tabular}
\medskip
\end{center}
\end{table}
\addtocounter{table}{-1}

\begin{table}
\caption{(continued)}
\footnotesize
\begin{center}
\medskip
\begin{tabular}{ccccrrrrrr} \hline\hline
(1) & (2) & (3) & (4) & (5) & (6) & (7) & (8) & (9) & (10)\\
Name      & $\tau_\nu$   & N$_\nu$ & $\sigma_{\rm S,N}$ & $S_{\rm int}$ & $S_m$ &  $\nu_m$ & $\alpha_{\rm fit}$ & $\gamma_{\rm min}$ & L$^{4-25\,GHz}_{\rm max}\,h^{-2}$  \\ \hline\hline
C5& 1984.2 & 3 & 12.5  & $34.7 \pm 2.3$   & $4.9\pm 1.1$  & $ 2.5\pm 0.4$ & -0.8 & 2.3 & $1.2\cdot 10^{43}$ \\
  & 1985.8 & 3 &  6.8  & $ 51.2 \pm 3.0$  & $3.3\pm 0.3$  & $ 6.0\pm 0.3$ & -0.6 & 3.3 & $6.1\cdot 10^{42}$\\
  & 1987.3 & 3 &  8.1  & $ 80.7 \pm 5.1$  & $5.1\pm 0.2$  & $ 7.0\pm 0.6$ & -0.8 & 4.3 & $5.4\cdot 10^{42}$ \\
  & 1988.2 & 4 &  7.5  & $ 98.0 \pm 4.0$  & $5.6\pm 0.5$  & $ 7.5\pm 0.4$ & -0.7 & 5.0 & $4.5\cdot 10^{42}$ \\
  & 1989.2 & 3 &  5.3  & $ 28.1 \pm 1.6$  & $4.7\pm 0.6$  & $ 2.5\pm 1.4$ & -0.9 & 5.7 & $1.7\cdot 10^{42}$ \\
  & 1990.3 & 3 &  5.4  & $ 26.8 \pm 1.1$  & $3.6\pm 0.1$  & $ 2.5\pm 1.2$ & -0.9 & 6.0 & $1.4\cdot 10^{42}$ \\
  & 1991.7 & 2 &  9.3  & $ 14.3 \pm 1.2$  & $(1.1)$       & $(3.2)$       & -0.5 & 5.7 & $6.6\cdot 10^{41}$ \\
  & 1992.5 & 3 & 12.3  & $  9.2 \pm 0.7$  & $1.5\pm 0.3$  & $ 2.2\pm 0.8$ & -0.9 & 5.0 & $7.4\cdot 10^{41}$ \\ 
  & 1993.8 & 2 &  5.7  & $  5.5 \pm 0.6$  & $(0.8)$       & $(2.3)$       & -0.8 & 2.8 & $1.4\cdot 10^{42}$ \\ \hline
\end{tabular}
\medskip
\end{center}
\end{table}
\addtocounter{table}{-1}

\begin{table}
\caption{(continued)}
\footnotesize
\begin{center}
\medskip
\begin{tabular}{ccccrrrrrr} \hline\hline
(1) & (2) & (3) & (4) & (5) & (6) & (7) & (8) & (9) & (10)\\
Name      & $\tau_\nu$   & N$_\nu$ & $\sigma_{\rm S,N}$ & $S_{\rm int}$ & $S_m$ &  $\nu_m$ & $\alpha_{\rm fit}$ & $\gamma_{\rm min}$ & L$^{4-25\,GHz}_{\rm max}\,h^{-2}$  \\ \hline\hline
C4& 1981.5 & 2 & 6.1   &$ 71.3 \pm 4.4$  & $(3.5)     $ & $(9.2)      $ & -0.2 & 3.2  & $7.4\cdot 10^{42}$ \\
  & 1982.4 & 2 & 6.7   & $162.7 \pm 7.1$ & $(8.0)     $ & $(10.0)     $ & -0.2 & 4.2  & $9.5\cdot 10^{42}$ \\
  & 1983.4 & 3 & 5.1   & $130.6 \pm 6.4$ & $7.5\pm 0.6$ & $11.6\pm 0.4$ & -0.5 & 4.7  & $5.6\cdot 10^{42}$ \\
  & 1984.2 & 2 & 5.4   & $113.2 \pm 6.2$ & $(5.6)     $ & $(14.1)     $ & -0.2 & 4.4  & $6.0\cdot 10^{42}$ \\
  & 1985.8 & 3&  6.4  & $ 36.3 \pm 3.0$  & $1.9\pm 0.1$ & $ 9.4\pm 0.6$ & -0.2 & 4.2  & $2.3\cdot 10^{42}$ \\
  & 1987.3 & 2 & 10.9  & $ 22.1 \pm 1.5$ & $(3.1)     $ & $(2.5)      $ & -0.8 & 5.5  & $1.5\cdot 10^{42}$ \\
  & 1988.2 & 3 & 17.2  & $ 12.2 \pm 0.8$ & $0.8\pm 0.1$ & $ 6.5\pm 1.4$ & -0.6 & 7.2  & $3.1\cdot 10^{41}$ \\
  & 1989.2 & 2 &  3.4  & $ 43.3 \pm 2.8$ & $(3.0)     $ & $(5.9)      $ & -0.8 & 9.6  & $7.5\cdot 10^{41}$ \\
  & 1990.3 & 2 &  22.0 & $  6.7 \pm 1.4$ & $(2.0)     $ & $(1.8)      $ & -1.4 & 12.2 & $1.3\cdot 10^{41}$ \\
  & 1991.7 & 2 &  19.3 & $  8.0 \pm 0.8$ & $(0.8)     $ & $(2.8)      $ & -0.6 & 14.9 & $5.7\cdot 10^{40}$ \\
  & 1992.5 & 3 &  22.1 & $ 15.7 \pm 1.2$ & $1.6\pm 0.1$ & $ 2.7\pm 1.6$ & -0.7 & 14.2 & $1.3\cdot 10^{41}$ \\ 
  & 1993.8 & 2 &  5.7  & $ 12.4 \pm 1.3$ & $(0.7)     $ & $(5.2)      $ & -0.1 & 9.8  & $1.7\cdot 10^{41}$ \\ \hline
\end{tabular}
\medskip
\end{center}
\end{table}
\addtocounter{table}{-1}

\begin{table}
\caption{(continued)}
\footnotesize
\begin{center}
\medskip
\begin{tabular}{ccccrrrrrr} \hline\hline
(1) & (2) & (3) & (4) & (5) & (6) & (7) & (8) & (9) & (10)\\
Name      & $\tau_\nu$   & N$_\nu$ & $\sigma_{\rm S,N}$ & $S_{\rm int}$ & $S_m$ &  $\nu_m$ & $\alpha_{\rm fit}$ & $\gamma_{\rm min}$ & L$^{4-25\,GHz}_{\rm max}\,h^{-2}$  \\ \hline\hline
C3& 1979.3 & 2 &  5.5  & $ 36.3 \pm 3.2$ & $(2.2)     $ & $(4.5)      $ & -0.2 & 7.3  & $9.3\cdot 10^{41}$ \\
  & 1980.6 & 2 &  5.8  & $ 19.9 \pm 1.1$ & $(3.2)     $ & $(2.3)      $ & -1.0 & 6.4  & $1.1\cdot 10^{42}$ \\
  & 1981.5 & 3 &  8.8  & $ 22.5 \pm 1.3$ & $2.4\pm 0.2$ & $ 2.7\pm 0.3$ & -0.8 & 6.0  & $1.1\cdot 10^{42}$ \\
  & 1983.4 & 2 & 12.9  & $ 13.1 \pm 0.7$ & $(2.0)     $ & $(2.4)      $ & -0.9 & 5.6  & $8.6\cdot 10^{41}$ \\
  & 1984.2 & 2 &  7.5  & $ 10.1 \pm 0.6$ & $(0.8)     $ & $(3.1)      $ & -0.5 & 5.7  & $4.9\cdot 10^{41}$ \\
  & 1985.8 & 2 & 29.5  & $  6.2 \pm 1.5$ & $(0.8)     $ & $(0.5)      $ & -0.5 & 6.2  & $5.8\cdot 10^{41}$ \\
  & 1988.2 & 2 & 15.7  & $  8.3 \pm 1.0$ & $(0.7)     $ & $(3.0)      $ & -0.5 & 7.9  & $2.1\cdot 10^{41}$ \\
  & 1989.2 & 2 &  6.9  & $ 17.1 \pm 1.2$ & $(1.0)     $ & $(16.8)     $ & -0.1 & 8.9  & $2.5\cdot 10^{41}$ \\
  & 1992.5 & 2 & 18.4  & $  8.2 \pm 1.2$ & $(0.4)     $ & $(8.4)      $ & -0.1 & 13.4 & $5.8\cdot 10^{40}$ \\
  & 1993.8 & 2 & 14.1  & $  3.1 \pm 1.4$ & $(0.3)     $ & $(2.5)      $ & -0.6 & 15.7 & $2.2\cdot 10^{40}$ \\ \hline
\end{tabular}
\medskip
\end{center}
\end{table}
\addtocounter{table}{-1}

\begin{table}
\caption{(continued)}
\footnotesize
\begin{center}
\medskip
\begin{tabular}{ccccrrrrrr} \hline\hline
(1) & (2) & (3) & (4) & (5) & (6) & (7) & (8) &(9) &(10)\\
Name      & $\tau_\nu$   & N$_\nu$ & $\sigma_{\rm S,N}$ & $S_{\rm int}$ & $S_m$ &  $\nu_m$ & $\alpha_{\rm fit}$ & $\gamma_{\rm min}$ & L$^{4-25\,GHz}_{\rm max}\,h^{-2}$  \\ \hline\hline
C2& 1979.3 & 2 & 34.3  & $ 13.1 \pm 0.9$ & $(0.9)     $ & $(3.5)     $ & -0.3 & 7.8  & $3.1\cdot 10^{41}$ \\
  & 1980.6 & 2 & 12.2  & $15.0 \pm 1.1$  & $(0.8)     $ & $(5.4)     $ & -0.1 & 8.2  & $2.9\cdot 10^{41}$ \\
  & 1981.5 & 3 & 10.9  & $16.9 \pm 1.0$  & $(2.0)     $ & $(1.4)     $ & -0.4 & 8.4  & $4.0\cdot 10^{41}$ \\
  & 1984.2 & 2 & 33.1  & $12.5 \pm 0.8$  & $(0.7)     $ & $(9.4)     $ & -0.2 & 9.1  & $1.9\cdot 10^{41}$ \\
  & 1989.2 & 2 &  7.5  & $10.7 \pm 1.2$  & $(0.6)     $ & $(5.4)     $ & -0.1 & 10.5 & $1.3\cdot 10^{41}$ \\
\hline
\end{tabular}
\end{center}
\tablecomments{The table columns are: 1 -- component name; 
2 -- multi frequency data set epoch; 3 -- number of frequencies
available for spectral fit; 4 -- mean fractional flux density error in
a spectral dataset [\%] ; 5 -- integrated 4--25\,GHz flux [$\cdot
10^{-14}$\,erg\,\,\,s$^{-1}$\,cm$^{-2}$]; 6 -- turnover flux density
[Jy]; lower limits are given in brackets; 7 -- turnover frequency [GHz];
upper limits are given in brackets; 8 -- fitted spectral index; 9 --
minimum Lorentz factor; 10 -- upper limit for 4--25\,GHz luminosity
[erg\,s$^{-1}$].}
\end{table}

\begin{table} []
\caption{Luminosity evolution in parsec--scale features of 3C345 
\label{tb-evolbet}}
\footnotesize
\begin{center}
\medskip
\begin{tabular}{ccc} \hline\hline
Component & $\log[L_{\rm 4-25\,GHz}(t)]$ & Comment \\ \hline\hline
D & $ 43.360 - 0.091\tau$ & 1981--88\\
C5 & $43.695 - 0.128\tau$ & \\
C4 & $43.517 - 0.182\tau$ &  \\
C3 & $42.280 - 0.110\tau$ & \\
C2 & $41.568 - 0.045\tau$ & \\
\hline
\end{tabular}
\end{center}
\tablecomments{ $\tau=t-1979.0\,$[years], where $t$ is current epoch.}
\end{table}

\begin{table} []
\caption{Correlation between the component luminosity and kinematics 
\label{tb-lum}}
\footnotesize
\begin{center}
\medskip
\begin{tabular}{cccc} \hline\hline
Component & $\log[L_{\rm 4-25\,GHz}(R_{\rm app})]$ & $\log[L_{\rm 4-25\,GHz}(R_j)]$ &
$\log[L_{\rm 4-25\,GHz}(R_j)]$ \\ 
 & & $\gamma =$const & $\gamma = \gamma_{\rm min}(t)$ \\ \hline\hline
C5 & $43.548 - 0.326\, R_{\rm app}$  & $43.323 - 0.654\, R_j$ & $43.534 - 0.668\, R_j$ \\
C4 & $43.346 - 0.661\, R_{\rm app}$  & $41.987 - 0.027\, R_j$ & $43.423 - 0.945\, R_j$ \\
C3 & $42.756 - 0.286\, R_{\rm app}$  & $41.221 - 0.018\, R_j$ & $42.609 - 0.567\, R_j$ \\
C2 & $41.932 - 0.100\, R_{\rm app}$  & $41.220 - 0.011\, R_j$ & $42.067 - 0.234\, R_j$ \\
All & $42.914 - 0.290\, R_{\rm app}$ & $42.060 - 0.030\, R_j$ & $43.066 - 0.638\, R_j$ \\
\hline
\end{tabular}
\medskip
\end{center}
\end{table}

\begin{table}
\caption{Component $S_m-\nu_m$ evolution in terms of shock--in--jet model \label{tb-SmNum}}
\footnotesize
\begin{center}
\medskip
\begin{tabular}{lcrrrl}\hline\hline
   & Period       & $\rho$ & $b$ & $\varepsilon$ & Stage \\ 
   &              & [$S_m\propto \nu_m^{\rho}$] & [$\delta \propto R^b$] & [$\nu_m\propto R^{\varepsilon}$] &  \\\hline\hline
 D:& \multicolumn{5}{c}{$a=1$, $s=1.2$} \\
   &              &        & $-0.5$ & $-1.1$ & Synchrotron \\
   & \raisebox{2ex}{1981.5--82.4} & \raisebox{2.0ex}{0.4} & $-$0.2 & $-$1.2 & Adiabatic   \\ \cline{2-6}
   & 1984.2--85.8 & 1.0    & $-0.9$ & $-1.4$ & Synchrotron \\
   & 1985.8--89.3 & 2.3    & $-$5.8 & $-$4.6 & Adiabatic   \\ \cline{2-6}
   & 1990.2--91.8 & 3.2    &  5.0 &  3.2 & Compton     \\
   & 1991.8--92.5 & $-$0.7   & $-$0.1 & $-$0.8 & Synchrotron \\
   & 1992.5--93.8 & 2.3    & $-$5.8 & $-$4.6 & Adiabatic   \\ \hline
C5:& \multicolumn{5}{c}{$a=1$, $s=2.4$} \\
   & 1985.8--88.3 & 5.1    &  2.0 &  1.0 & Compton     \\
   & 1988.3--89.2 & 0.0    & 0.0 & $-$1.0 & Synchrotron \\
   & 1989.2--92.5 & 1.8    & $-$2.6 & $-$3.1 & Adiabatic \\  
   &              & (0.8)  & ($-$0.2) & ($-$1.5) & Adiabatic\tablenotemark{1}\\ \hline
C4:& \multicolumn{5}{c}{$a=1$, $s=2.4$} \\
   & 1983.4--88.2 & 2.8    & $-$9.7 & $-$7.8 & Synchrotron \\
   & 1988.2--92.5 & $-$0.9   &  0.8 & $-$0.8 & Adiabatic   \\
\hline
\end{tabular}
\end{center}
\tablenotetext{1}{ used for the model described in section \ref{sc:c5-spkin}}
\end{table}

\begin{table}[]

\caption{Kinematics of C5 in the rest frame\label{tb-c5stages}}
\footnotesize
\begin{center}
\medskip
\begin{tabular}{cccccc} \hline\hline
 $t_1-t_2$ & $R_2/R_1$ & $\Delta R$ & $R_1$ & $\Delta R$ & $R_1$ \\
$[{\rm years}]$ &  & [pc] & [pc] & [pc] & [pc] \\ \cline{3-6}
  &  & \multicolumn{2}{c}{($\delta_{01} = 1.05)$} & 
    \multicolumn{2}{c}{($\delta_{02} = 2.80)$} \\ \hline
1985.8--88.2 & 1.1 & 4.2 & 29.3 & 6.1 & 42.0 \\
1988.2--89.2 & 2.3 & 5.6 & 4.4 & 3.9 & 3.0 \\
1989.2--92.5 & 1.6 & 8.3 & 13.4 & 14.5 & 23.2 \\
\hline
\end{tabular}
\medskip
\end{center}
\end{table}

\begin{table}
\caption{Parameters of the flares in 3C\,345}
\label{tb:flares}
\footnotesize
\begin{center}
\medskip
\begin{tabular}{cccccccc}\hline\hline
Flare & \multicolumn{3}{c}{Spectrum} & &\multicolumn{3}{c}{Light Curve} \\\cline{2-4} \cline{6-8}
      & $t_{\rm flare}$ & $N_{\rm rel}$ & $\tau_{\rm flare}$ && $t_{\rm flare}$ & $N_{\rm rel}$ & $\tau_{\rm flare}$ \\
1 & (1981.4)  & $(1.0\cdot 10^3)$ & (0.5)  && 1981.4 & $1.2\cdot 10^4$ & 0.4 \\
2 & 1984.6  & $1.5\cdot 10^3$ & 0.3  && 1984.2 & $1.5\cdot 10^4$ & 0.3 \\
3 & 1987.8  & $3.3\cdot 10^1$ & 0.2  && 1986.6 & $7.3\cdot 10^1$ & 0.2 \\
4 & 1992.0  & $1.8\cdot 10^3$ & 0.4  && 1991.7 & $3.0\cdot 10^4$ & 0.4 \\
5 & (1995.2) & $(1.8\cdot 10^3)$ & (0.4)  && 1995.2 & $5.0\cdot 10^2$ & 0.2 \\
6 &         &                 &      && (1999.4) & $(3.0\cdot10^4)$& (0.4) \\
\hline
\end{tabular}
\end{center}
\tablecomments{$t_{\rm flare}$ -- epoch of the flare; $N_{\rm rel}$ --
relative density increase; $\tau_{\rm flare}$ -- duration of the flare [yrs].
Bracketed values refer to the fits which are insufficiently constrained by
the data.}
\end{table}

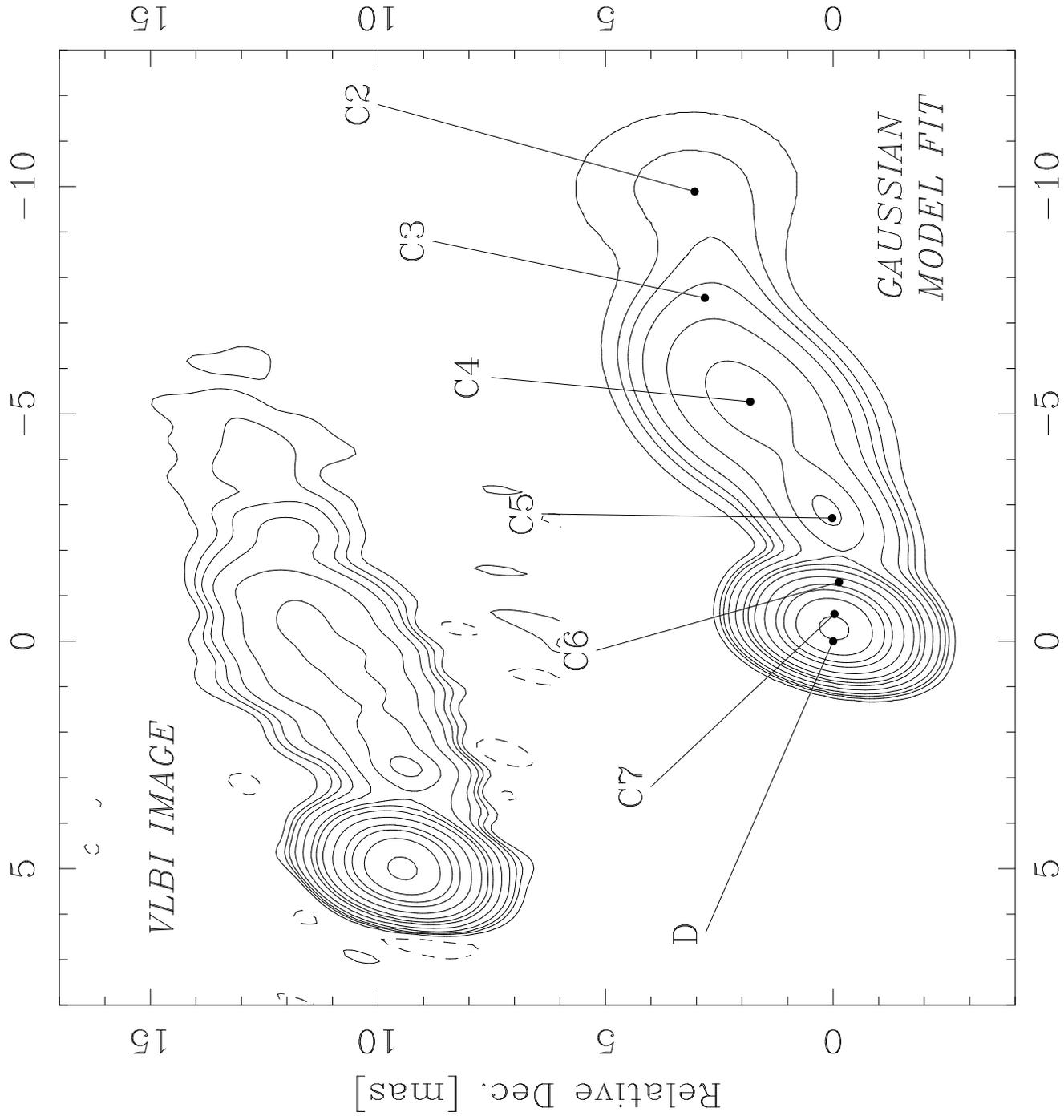
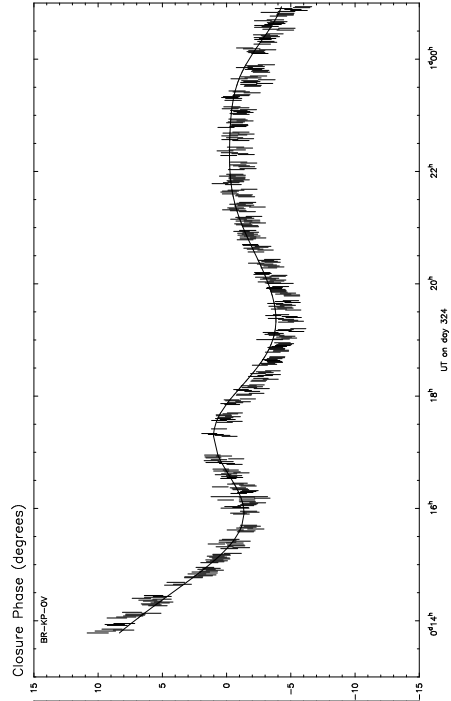
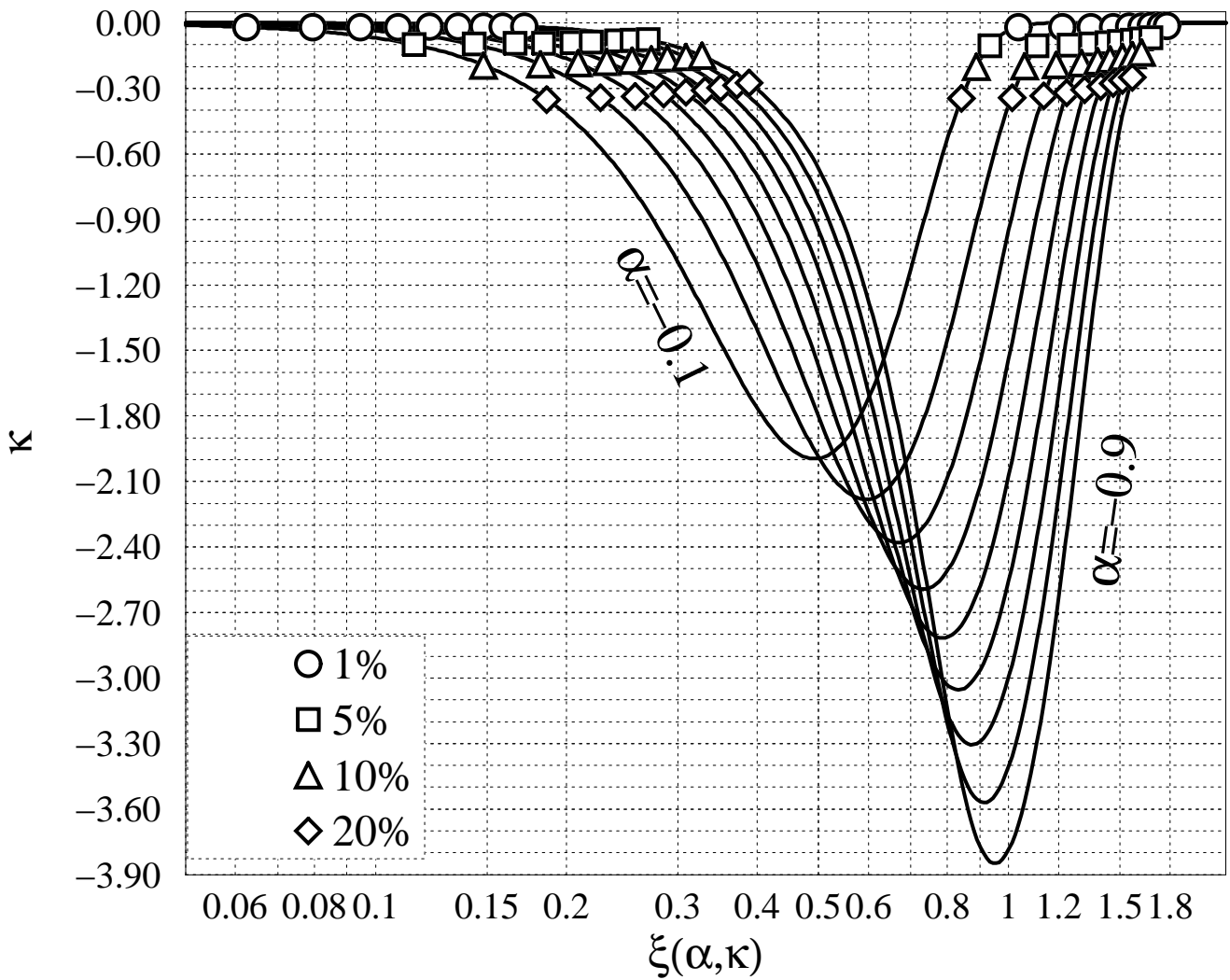
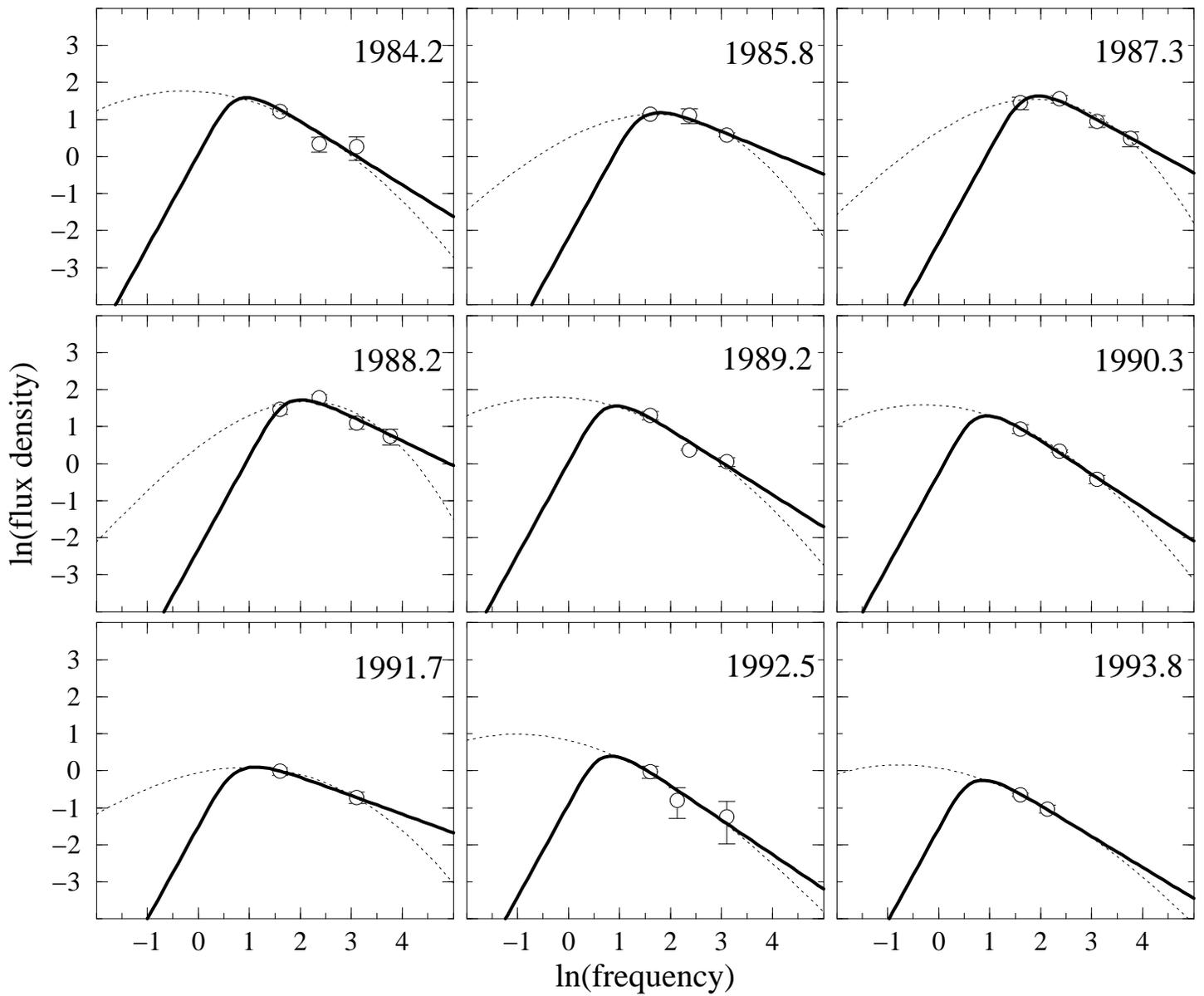
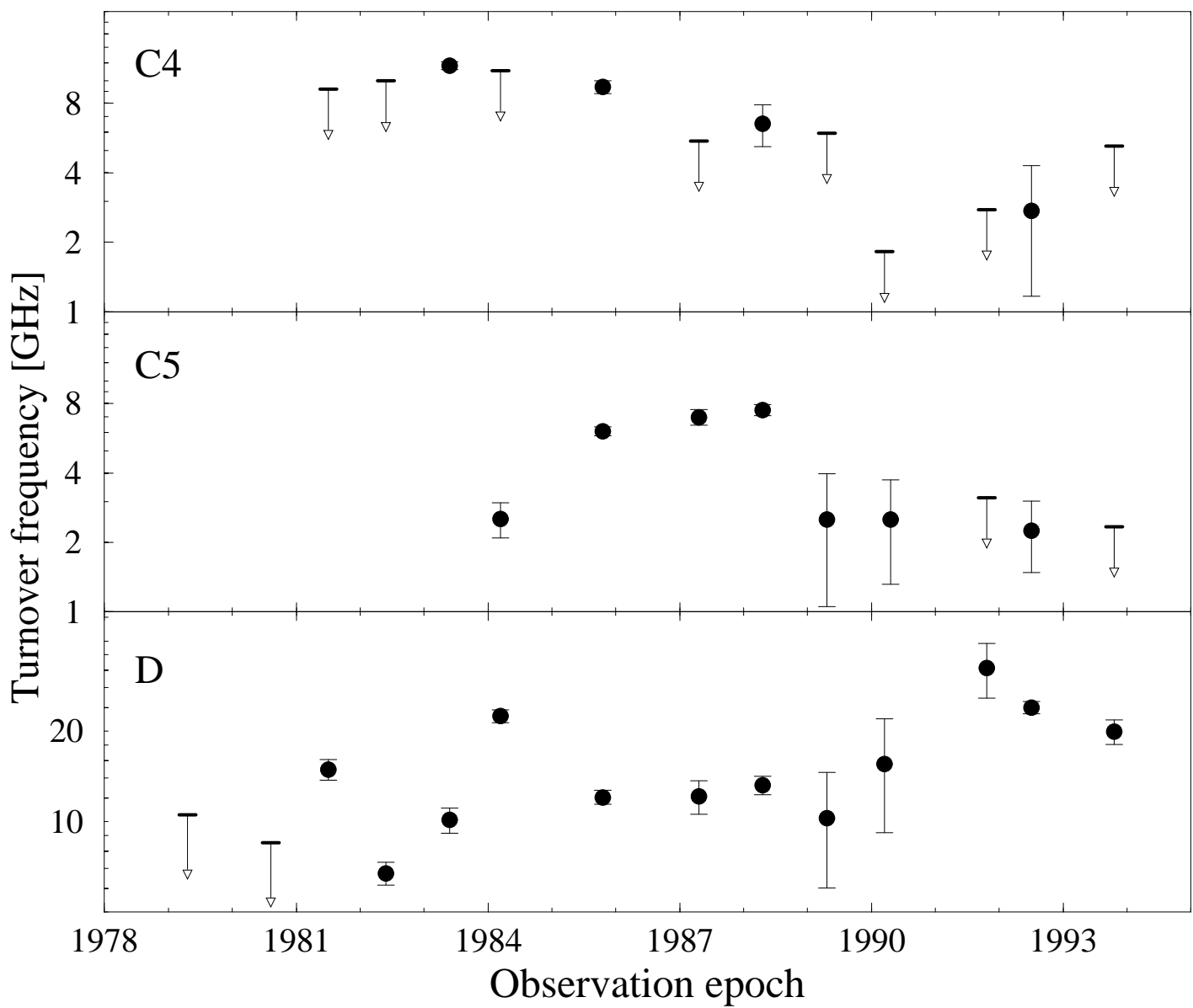
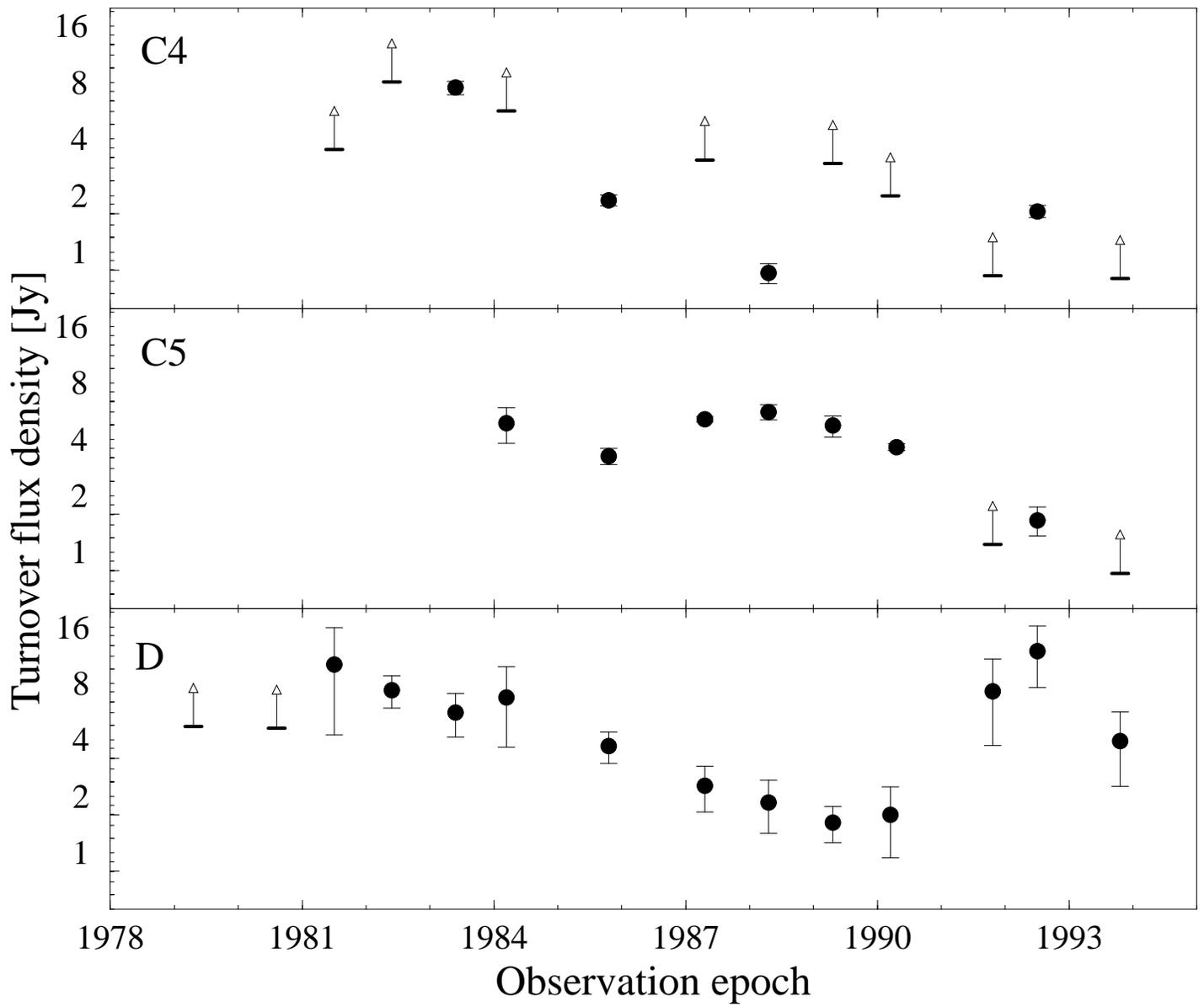
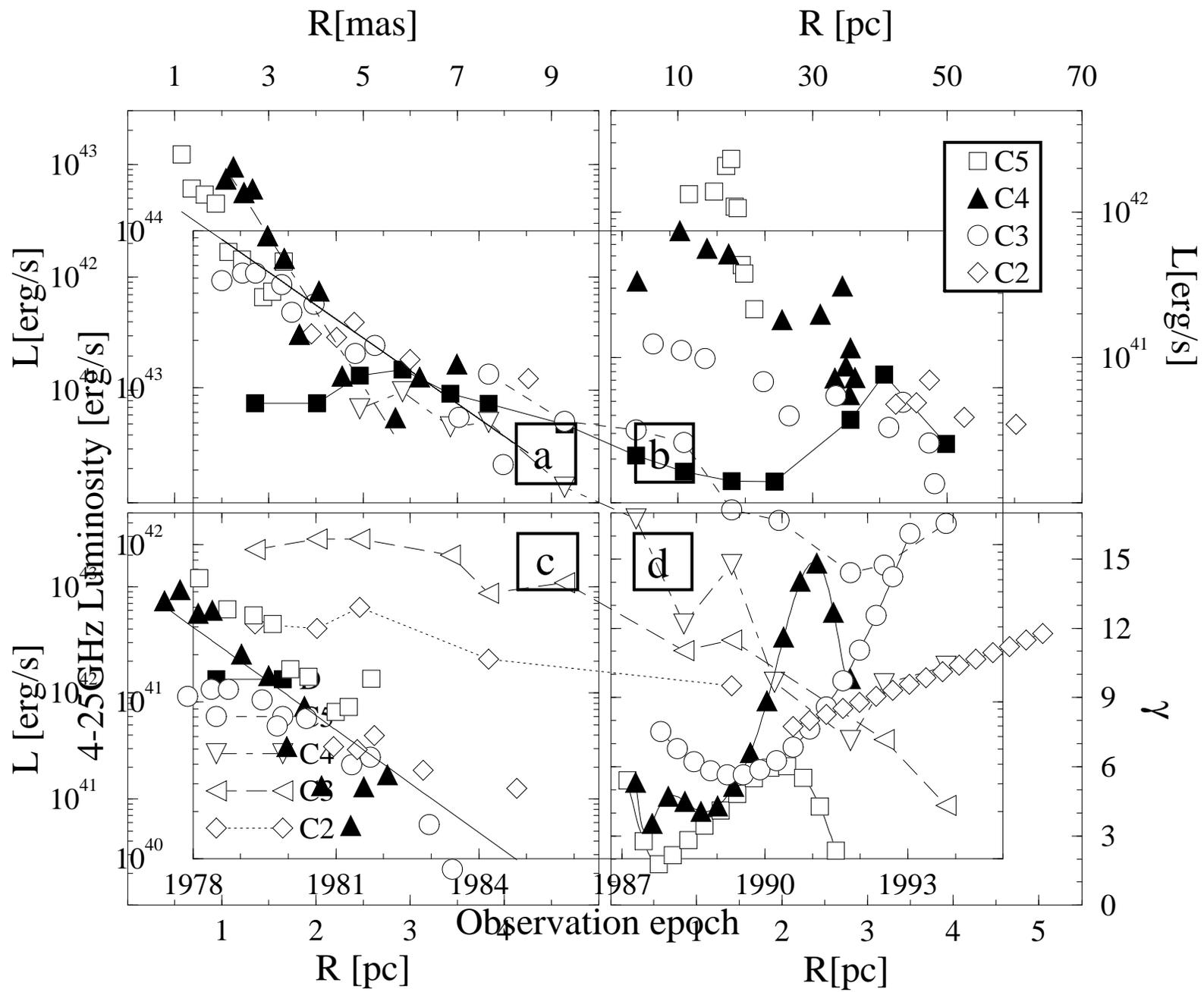
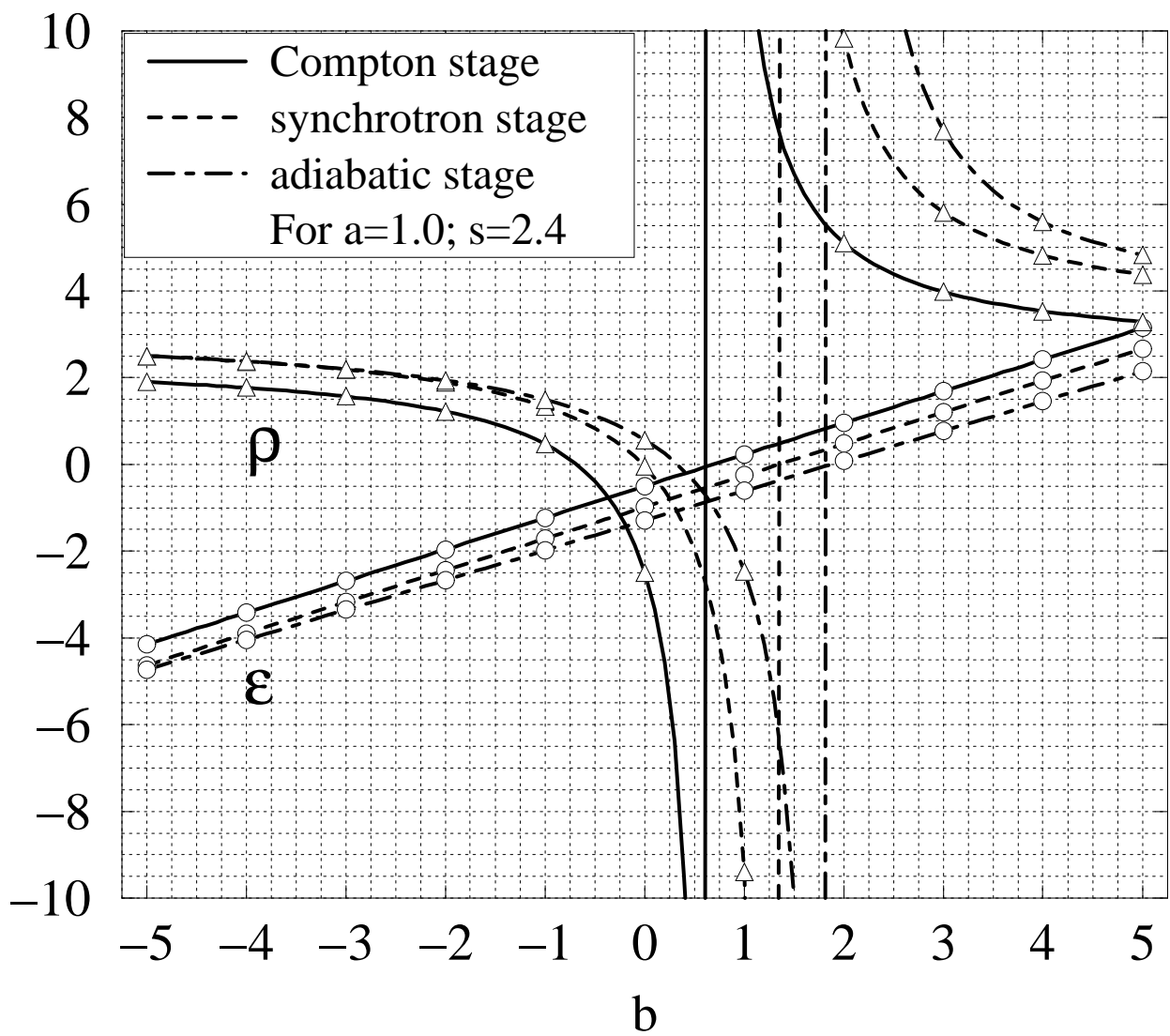
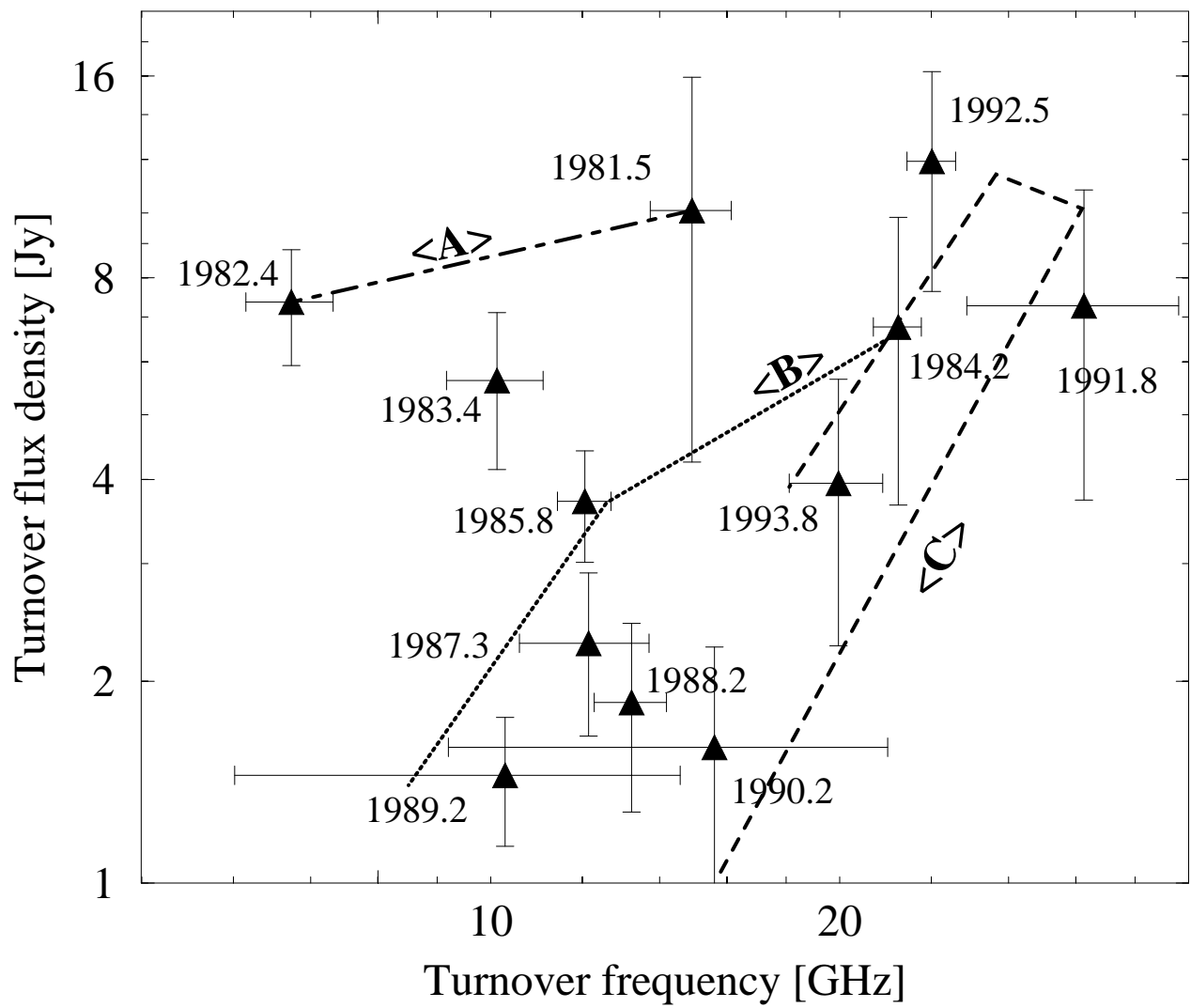
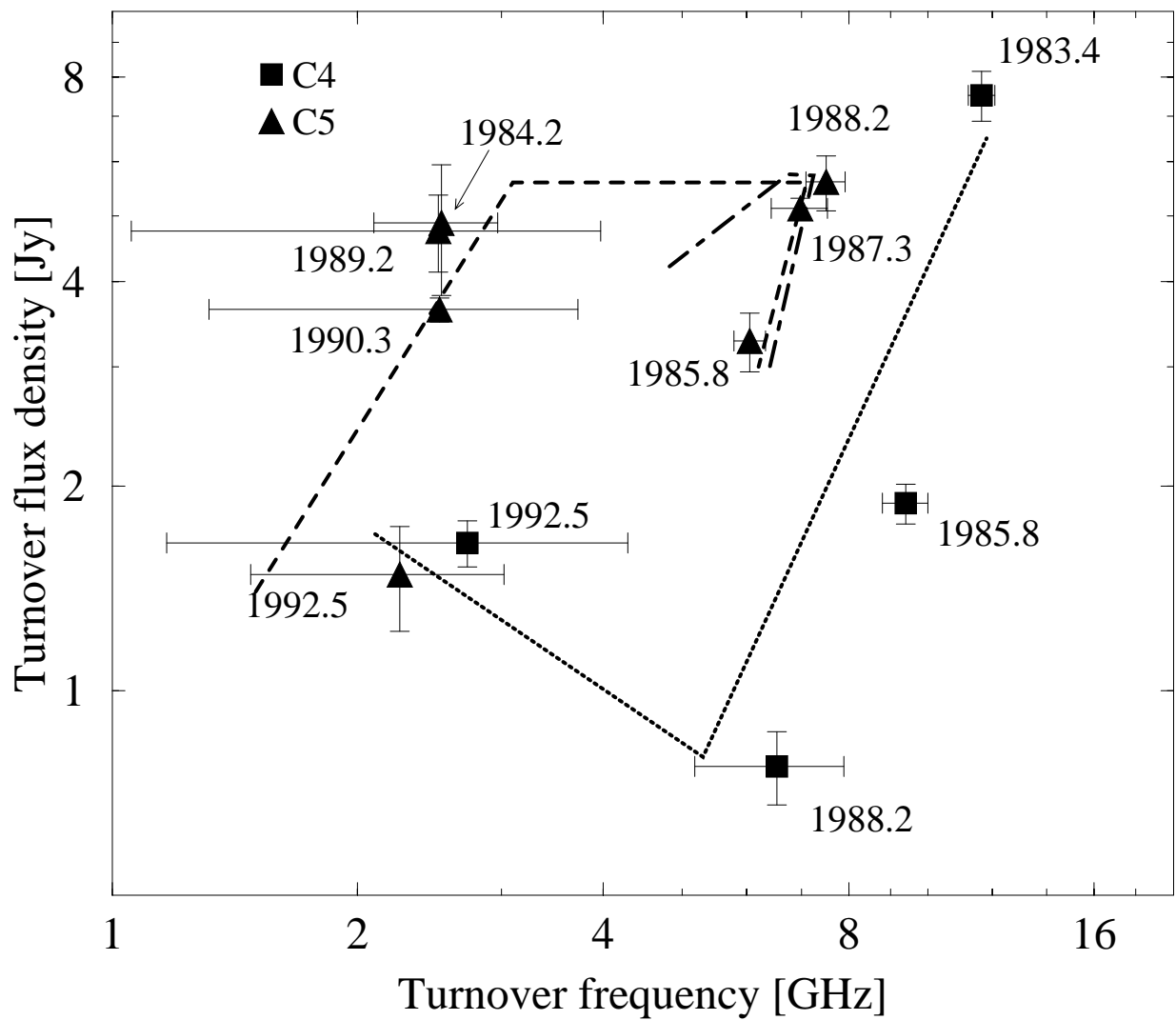
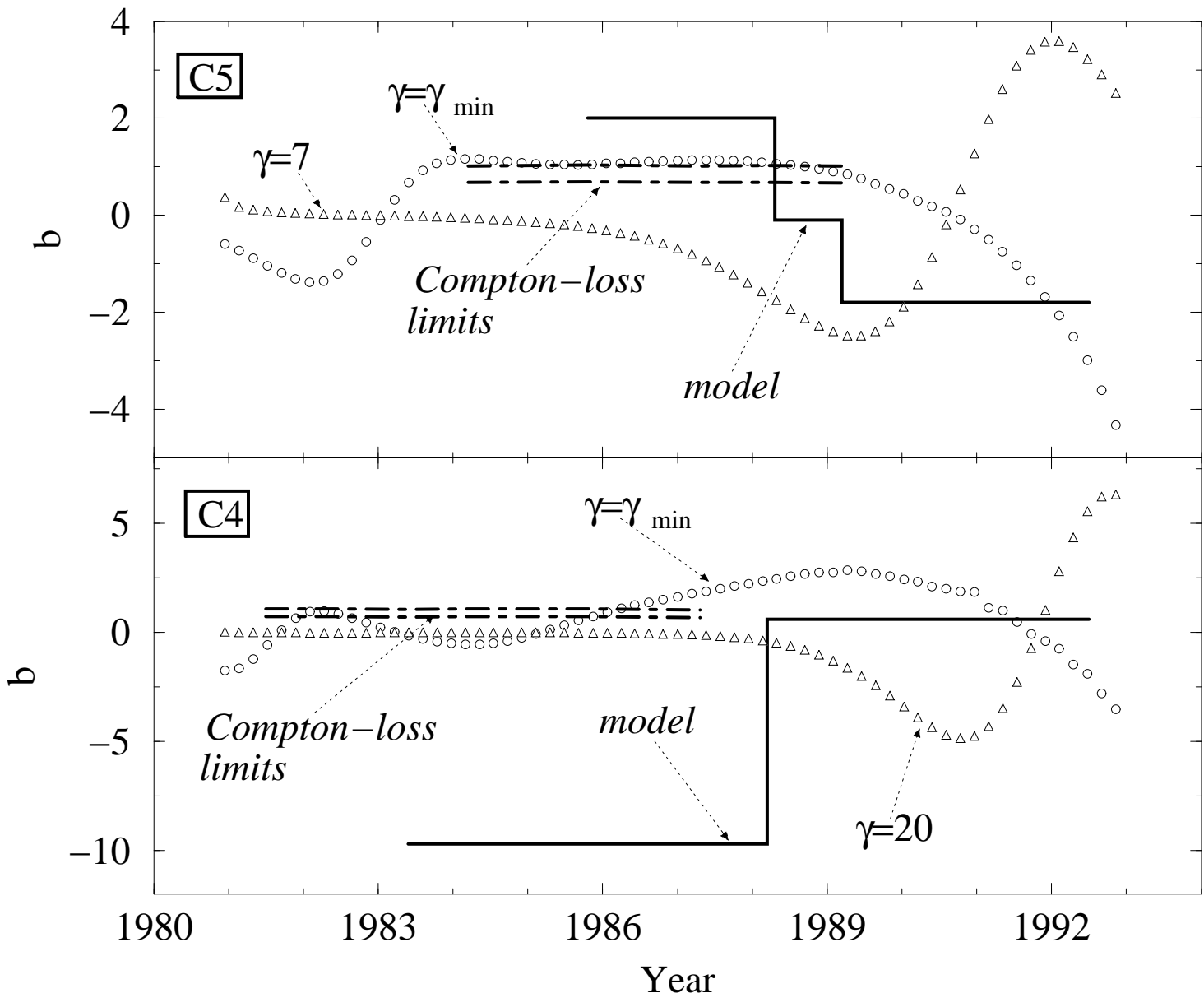
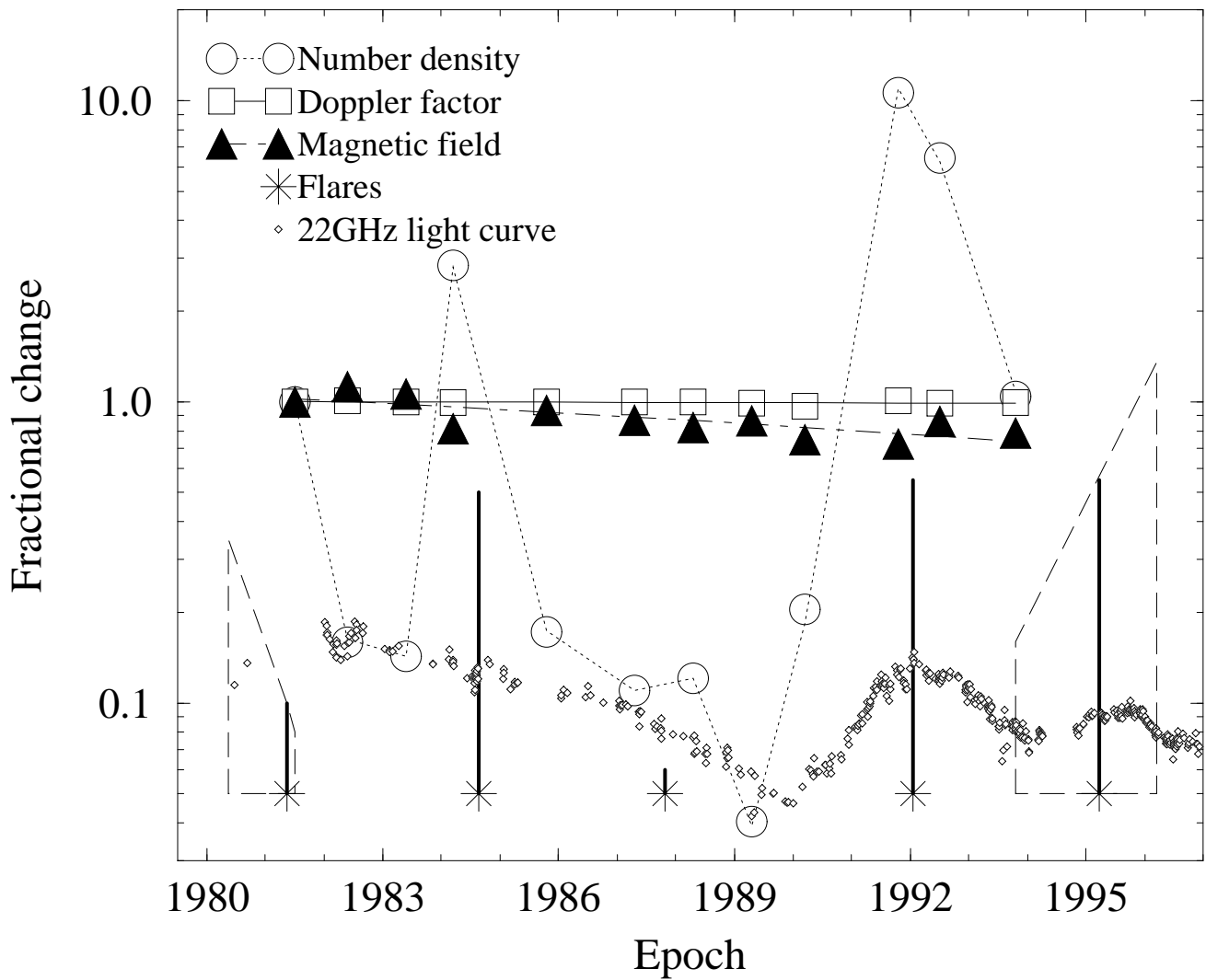
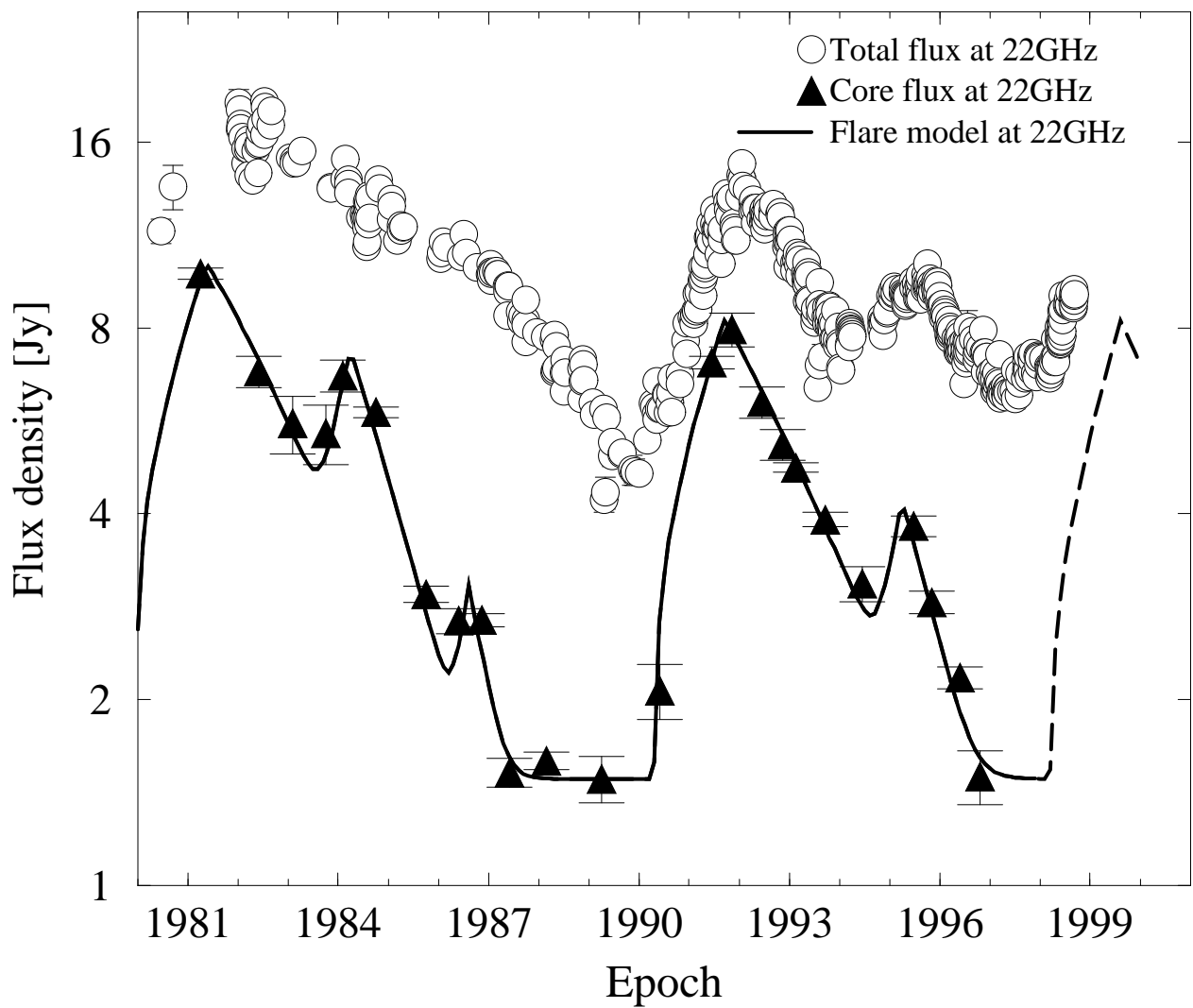
\end{document}